\documentclass[fleqn,usenatbib,twocolumn,useAMS]{mnras}

\usepackage{mathptmx}
\usepackage{txfonts}

\usepackage[T1]{fontenc}
\usepackage{ae,aecompl}
\usepackage[utf8]{inputenc}
\usepackage{longtable}
\usepackage{supertabular}
\usepackage{multicol}
\usepackage{multirow}

\usepackage{graphicx}	

\usepackage{etoolbox}
\makeatletter
\patchcmd\@combinedblfloats{\box\@outputbox}{\unvbox\@outputbox}{}{\errmessage{\noexpand patch failed}}
\makeatother

\newcommand{\ms}{M$_{\odot}$}

\title[A new assessment of solar s- and r- components]{Chemical evolution with rotating massive star yields \\
II. A new assessment of the solar s- and r- process components }

\author[Prantzos et al.]{
N. Prantzos,$^{1}$\thanks{E-mail: prantzos@iap.fr}
C. Abia,$^{2}$
S. Cristallo$^{3,4}$
M. Limongi,$^{5,6}$
A. Chieffi$^{7,8}$
\\
$^{1}$Institut d'Astrophysique de Paris, UMR7095 CNRS, Sorbonne Université, 98bis Bd. Arago, 75104 Paris, France\\
$^{2}$Departmento de F\'\i sica Te\'orica y del Cosmos, Universidad de Granada, E-18071 Granada, Spain\\
$^{3}$
Istituto Nazionale di Astrofisica - Osservatorio Astronomico d'Abruzzo, Via Maggini snc, I-64100, Teramo, Italy\\
$^{4}$
Istituto Nazionale di Fisica Nucleare - Sezione di Perugia, Via Pascoli, I-06123, Perugia, Italy\\
$^{5}$
Istituto Nazionale di Astrofisica - Osservatorio Astronomico di Roma, Via Frascati 33, I-00040, Monteporzio Catone, Italy\\
$^{6}$
Kavli Institute for the Physics and Mathematics of the Universe, Todai Institutes for Advanced Study, the University of Tokyo, \\
  Kashiwa, Japan 277-8583 (Kavli IPMU, WPI)\\
  $^{7}$
Istituto di Astrofisica e Planetologia Spaziali, INAF, via Fosso del cavaliere 100, 00133 Roma - Italy \\
$^{8}$
Monash Centre for Astrophysics (MoCA), School of Mathematical Sciences, Monash University, Victoria 3800, Australia\\  
}

\date{Accepted XXX. Received YYY; in original form ZZZ}

\pubyear{2019}
\begin{document}
\label{firstpage}
\pagerange{\pageref{firstpage}--\pageref{lastpage}}
\maketitle

\begin{abstract}
The decomposition of the Solar system abundances of heavy
isotopes  into their s- and r- components plays a key role in our understanding of the corresponding nuclear processes and the physics and evolution of their astrophysical sites. We present a new method for determining the s- and r- components of the Solar system abundances, fully consistent with our current understanding
of stellar nucleosynthesis and galactic chemical evolution. The method is based on a  study of the  evolution of the solar neighborhood with a state-of-the-art 1-zone model, using recent yields of low and intermediate mass stars as well as of massive rotating stars.  
We compare our results with previous studies and we provide tables with the isotopic and elemental contributions of the s- and r-processes to the Solar system composition. 

\end{abstract}



\section{Introduction }
\label{sec:Introduction}

In their  compilation and analysis of Solar system isotopic abundances \cite{Suess1956} were the first to notice that, if heavier than Fe nuclei are formed by successive capture of neutrons, one should expect two abundance peaks for each of the regions near magic neutron numbers: a sharp one  at the position of the magic nucleus, from material pilled up there due to the low neutron capture cross-section when neutron captures take place near the $\beta$-stability valley;  and a smoothed one  at a few mass units below, from material made by neutron captures occurring in the neutron-rich side of the stability valley and radioactively decaying after the end of the process.

Building on that compilation, \cite{bur57} worked out the details of the two nucleosynthetic processes, which they called s- and r-, respectively\footnote{There are observational indications of intermediate density neutron capture processes (i.e. between the s- and r- process), like the i- process \citep{cow77,dar14,ham16}, possibly occurring in rapidly accreting white dwarfs \citep{deni17}, proton ingestion episodes in low-metallicity low-mass asymptotic giant branch (AGB) stars \citep{cri16} or super-AGB stars \citep{jon16}.}.  The former ({\it slow}) would occur on timescales long with respect to the lifetimes of radioactive nuclei along the neutron path, i.e. tens to thousands of years, as a result of  low neutron densities  $N_n\sim$10$^6\div 10^7$ cm$^{-3}$. The latter ({\it rapid}) would take place on short timescales of the order of 1 s, as a result of high neutron densities $N_n>10^{24}$  cm$^{-3}$. \cite{bur57} also  noticed that, along the s- process path (i.e. the valley of nuclear stability), the product of the neutron capture cross-section $\rm \sigma_A$ and the abundance N$_A$ of a nucleus with mass number A$>70$ is a smooth function of A, first declining up to A$\sim$100 and then levelling off up to A$=208$. They attributed that feature to the operation of the s-process in two different regimes, the former one having "not enough neutrons available per $^{56}$Fe nucleus to build the nuclei to their saturation abundances", while the constancy of $\rm \sigma_A N_A$ in the latter is "strongly suggestive of steady flow being achieved and of all of the nuclei reaching their saturation abundances".

Following the work of \cite{Weigert1966}, the environment provided by low and intermediate mass stars (LIMS) on their AGB phase was identified by \cite{SchwarHarm1967} and \cite{Sanders1967} as a promising site for the operation of the s-process.  Today, those stars are thought to produce the bulk of the s-isotopes above A$\sim90$ during their thermal pulses, with neutrons released mainly by the $^{13}$C($\alpha$,n)$^{16}$O reaction (see \citealt{str95,Gal98} and references therein). On the other hand, \cite{Peters1968} suggested that in the He-burning cores of massive stars, neutrons released by the $^{22}$Ne($\alpha$,n)$^{25}$Mg reaction should also produce s-nuclei. Today, those stars are thought to produce the s-nuclei in the regime of "few neutrons per $^{56}$Fe seed", i.e. below A$\sim90$ \citep{Couch1974,Lamb1977,Busso1985}: stellar models - including those of \cite{Prantzos1987} with mass loss - show that despite the large abundance of $^{22}$Ne, most of the released neutrons are captured by its progeny $^{25}$Mg and other abundant nuclei, leaving few neutrons to be captured by $^{56}$Fe 
\cite[see][for details of the "neutron economy trio", i.e. the roles of neutron sources, seed and poisons as function of metallicity in the case of massive stars]{Pra90}. In contrast, in the  thermally pulsing phase of AGBs, the periodic mixing of protons in the He-layer maintains the $^{13}$C source to a high abundance level - through $^{12}$C(p,$\gamma$)$^{13}$C -  and releases sufficient neutrons to reach the "saturation regime". Thus, both the mechanism(s) and site(s) of the s- process are considered to be sufficiently well known \cite[see e.g.][and references therein]{Kap11}.

On the other hand, the situation with the site of the 
r-process is still unsatisfactory.
After more than fifty years of research on its astrophysical origin(s), the identification of a fully convincing  site remains still elusive. An exhaustive description and discussion of experimental, observational and theoretical aspects of the r-process, as well as on the sites so far proposed is provided in the recent reviews of \cite{Cowan2019}  and \citet{thi17}. However, up to date, no numerical simulation in the proposed scenarios has been able to fully reproduce the observed distribution of the r-process elemental and isotopic abundances in the Solar system. Nowadays the neutron star merging (NSM) scenario is given support by the recent joint detection of electromagnetic and gravitational signal from the
$\gamma-$ray burst GW170817/GRB170817A \citep[see][and references therein]{Pian2017}, 
and, in particular, by the identification of the neutron-capture element Sr in the spectrum of the associated kilonova  AT2017gfo \citep{Watson2019}.
However,  it is not yet completely understood which component of those systems (dynamical, disk, $\nu$-wind) dominates the nucleosynthesis, since any one of them may cover a wide range of chemical distributions, depending on the adopted input parameters \citep{Ro15,fm16,wu16,pe17}. An additional important source of uncertainty comes from the nuclear inputs adopted to calculate the r-process nucleosynthesis, the most important ones being nuclear masses, $\beta$-decay rates and nuclear fission models \citep{ei15,thi17}. Finally, the observed evolution of the  r-elements in the Galaxy is hard (albeit not impossible) to conciliate with our current understanding of the occurrence rate of NSMs, regarding both the early (halo) and the late (disk)  phases of the Milky Way \citep{Tsuji2014,Ishimaru2015,Ojima2018,cot18,Hotokezaka2018,cot18,Guiglion2018,Wehmeyer2019,siegel2019,Haynes2019,Cote2019}.

The decomposition of the Solar system abundances of heavy elements into their s- and r- components, played and will continue to play a pivotal role in our understanding of the underlying nuclear processes and the physics and evolution of the corresponding sites. The s-contribution can be more easily determined, since isotopes dominated by the s-process form close
the $\beta-$stability valley. Their nuclear properties ($\beta-$decay half-times, nuclear cross sections, etc.) are more easily measured, while the astrophysical sites are better understood today. On the other hand, due to the large astrophysics and nuclear physics uncertainties related with the r-process, its contribution
to the isotopic solar abundances has been so far  deduced by a simple subtraction of the s-process contribution from the observed solar value.

In this work, we present a new method for determining the s- and r-components of the Solar system abundances. It is based on a global study of the  evolution of the solar neighborhood with a state-of-the-art 1-zone model of galactic chemical evolution (GCE), which is presented in detail in \cite{pra18} - Paper I hereafter - and adopts recent stellar yields of rotating massive stars (from \citealt{lim18}) and of LIM stars (from \citealt{Cr15}).

The plan of the paper is as follows: In \S \ref{sec:Methods}, we review the various methods used so far in order to derive the s-component of the isotopic abundances of the heavy nuclei, and we discuss their shortcomings. In \S \ref{sec:OurMethod}, we present in detail our new method and its assumptions. In \S \ref{sec:Results}   we present our results. We compare first the isotopic contributions to previous studies (\S \ref{subsec:S-fractions}) , as well as to the measured Solar system abundances taking into account the uncertainties of the latter (\S \ref{subsec:Comparison_SS}). We discuss the resulting $\rm \sigma_A N_A$ curve in \S \ref{subsec:sigmaN} and we derive the r-residuals in \S \ref{subsec:R-component}. In \S \ref{subsec:Elemental_anundances} we derive the elemental s- and r-components, and finally in \S\ref{sec:Summary} we summarized the main results of this study.

\section{Determination of s- and r- abundances}
\label{sec:Methods}

The "classical" (or "canonical") s-process model was originally
proposed by \citet{bur57} and developed by \citet{Cla67}. In this model
two main assumptions are made: a) the s-process temperature is constant, allowing  one to adopt well determined neutron-capture cross sections; b) nuclei on the s-process path are either stable ($\tau_{\beta}>>\tau_n$) or sufficiently short-lived that the neutron capture chain continues with the daughter nucleus ($\tau_{\beta}<<\tau_n)$. This second assumption, however, is not valid at the s-process branchings ($\tau_{\beta}\sim \tau_n$), which requires a special treatment \citep[see e.g.][]{kap89}. In addition, the classical model assumes that some stellar material composed by iron nuclei only is exposed to the superposition of 3 exponential distributions of the time-integrated neutron exposure, defined as $\tau_o=\int_o^tN_nv_Tdt$ (where $v_T$ is the thermal neutron velocity at the temperature T). The 3 exponential distributions are usually referred to as the "weak" component (responsible of the production of the $70\leq A\leq 90$ s-nuclei), the "main" component (for the $90\leq A \leq 204$ isotopes) and the "strong component" (for $A > 204$). For long-enough exposures, the equations governing the evolution of the s-nuclei abundances result in equilibrium between the production and destruction terms, leading to a constant product, $\sigma_AN_A$, of neutron cross section and s-process abundance. Although this condition is not completely reached, the product $\sigma_AN_A$ shows a very smooth dependence on mass number (see, e.g., \citealt{cla68}). Therefore, the product $\sigma_AN_A$ for a given isotope is fully determined  by the cross section, after the parameters $\tau_o$ and the number of neutrons captured per $^{56}$Fe seed nucleus are fixed. The goal of the classical approach is to fix the empirical $\sigma_AN_A$ values {\it for the s-only isotopes}, i.e. nuclei that are shielded against the r-process by the corresponding stable isobar with charge $Z-1$ or $Z-2$ (see  \S \ref{subsec:S-R-only} for a discussion about our selection of s-only isotopes). Once the Solar system s-only distribution is fitted, the s-contribution for the rest of the "mixed" isotopes (with both a s- and r-contribution)  are automatically obtained. Finally, the r-contribution is derived just subtracting this s-contribution N$_{s,A}$ from the measured total Solar system abundance N$_{A}$.  This classical method has been used frequently in the literature,  providing satisfactory results as the measurement of neutron cross sections have been improving  during the years \citep[see e.g.][]{kap89,sne08,Kap11}. 

However, the classical model is affected not only by observational and nuclear input data uncertainties, but also by the assumption that the s-process operates at a fixed constant  temperature and neutron and electron
density, and by the hypothesis that the irradiation can be considered as exponential one. 
To test the influence of these assumptions, \citet{gor99} \citep[see also][]{arn07} developed the so-called "multi-event" s-process, which constitutes a step forward in the canonical method. The multi-event approach assumes a superposition of a number of canonical events taken place in different thermodynamic conditions, namely: a temperature range $1.5 \leq 
\rm{T(K)}/10^8 \leq 4$, neutron densities $7.5\leq \rm{log~N_n(cm^{-3})}\leq 10$ and a unique electron density $\rm{N_e}=10^{27}$ cm$^{-3}$. Each canonical event is characterized by a given neutron irradiation on the $^{56}$Fe seed nuclei during a given time at a constant temperature and neutron density.  These conditions try to mimic the astrophysical conditions characterizing the site of the s-process, although it is well known that temperature and neutron density are not constant during the s-process  \citep[see e.g.][and references therein]{Kap11}. The s-only nuclei abundance distribution obtained with that method is remarkably close to the solar observed one, because of the minimization procedure adopted in the selection of the aforementioned parameters. However, it presents non-negligible deviations from the classical method in the regions $A\leq 90$ and A$\geq 204$, mainly because the resulting neutron exposures in the multi-event model clearly deviate from exponentials. 
Within the multi-event model it was possible to evaluate the major uncertainties (both nuclear and due to abundance measurements) affecting the prediction of the s-(r-)abundance distribution.  \citet{gor99} concluded that the uncertainties in the observed meteoritic abundances and the relevant $(n,\gamma)$ rates have a significant impact on the predicted s-component of the solar abundance and, consequently, on the derived r-abundances, especially concerning the s-dominated nuclei (see also \citealt{nishi17} and \citealt{cescu18}).

Stellar models  of LIM stars during the AGB phase and of massive stars during hydrostatic core He-burning and shell C-burning (the two widely recognized sites of the s-process), have shown that the interplay of the different thermal conditions for the $^{13}$C and $^{22}$Ne neutron sources is hardly represented by a single set of effective parameters constant in time \citep{Bus99,str06,lim18}, such as those used in the classical (or the multi-event) approach. In an effort to overcome this shortcoming,  the results of the "{\it stellar}" model have been used to estimate the contributions of the s- and r-process to the Solar system abundances. This method is based on post-processing nucleosynthesis calculation performed 
in the framework of  "realistic"  stellar models.  The first attempt  to apply this method was made by \citet{Gal98} and \citet{Ar99}, and more recently by \citet{bis10} with updated nuclear input. These authors showed that the solar s-process main component can be
reasonably reproduced by a post-processing calculation from a particular choice (mass and extension) of the $^{13}$C pocket (the main neutron source in AGB stars) by averaging the results of stellar AGB models \citep{Gal98} between 1.5 and 3 M$_\odot$ with [Fe/H]$\sim -0.3$. This model is particularly successful in reproducing the s-only nuclei solar abundances and showed general improvements with respect to the classical method, especially in the
mass region $A< 88$ \citep{Ar99}. In fact, all these nuclei (mainly produced by the weak s-component) are synthesized in much smaller quantities. This difference is caused by the very high neutron exposures reached in the stellar model, which favor the production of heavier elements. In particular, at the s-termination path, $^{208}$Pb is produced four times more than in the classical approach. Nevertheless, the stellar model  used to derive the physical inputs of post-process calculations are affected by several theoretical uncertainties. One of the less constrained physical mechanisms is the one leading to the formation of the $^{13}$C pocket, which forms at the base of the convective envelope after each TDU episode. Different processes have been proposed as responsible of the formation of such a pocket:  convective overshoot \citep{he97}, gravity waves \citep{denito,uba}, opacity induced overshoot \cite{Cr09} and mixing induced by magnetic mixing \citep{Tr16}. Other critical quantities are the mass fraction dredged-up after each thermal instability (third dredge up, TDU) during the AGB phase, and the mass-loss rate.  Actually, the two processes are degenerate, since  the number (and the efficiency) of TDUs is determined by the  mass of the H-exhausted core and of the H-rich convective envelope, which in turn depend on the adopted  mass-loss rate. However, AGB stellar models show that an asymptotic s-process distribution is reached after a limited number of pulses, so the mass-loss uncertainty mainly affects the total yield of the s-processed material, and not so much the shape of the resulting distribution (see e.g. Fig. 12 in \citealt{Cr15b}). In the "{\it stellar}" model, the r-residuals are calculated subtracting the arithmetic average of the 1.5 and 3 M$_\odot$ models at [Fe/H]$\sim -0.3$ ($Z\sim 1/2 Z_\odot)$\footnote{We adopt here the usual notation [X/H]$=$ log (X/H)$_\star-$ log (X/H)$_\odot$, where (X/H)$_\star$ is the abundance by number of the element X in the corresponding object.} best reproducing the main s-component to the observed solar abundances. In \citet{Ar99} the s- and r-components obtained by the stellar model method are compared to the classical one for nuclei $A>88$, together with the corresponding uncertainty determined from the cross sections and solar abundances. Uncertainties in the s- and r-residuals coming from the stellar model itself are, however,  difficult to estimate.

The massive star contribution to the solar s- only composition has been explored with non-rotating stellar models in e.g. \cite{Pra90,Rai93} and more recently, with rotating massive stars in  \cite{Pignatari2008,Fri16,Choplin2017,cho18,lim18}. Such models have their own  uncertainties (mass loss, mixing, nuclear etc.). The role of rotation, in particular, is poorly explored and understood at present.  The main reason is that the rotation driven instabilities are included in a parametric way, and this means that the efficiency with which fresh protons are ingested in the 
He-burning zone is not based on first principles but it is determined by two free parameters that must be calibrated. The calibration adopted in the models adopted in this paper is discussed in detail in \cite{lim18}. Moreover, since the proton ingestion scales directly with the initial rotational velocity (and hence the neutron flux as well), the adopted initial distribution of rotational velocities (IDROV) plays a pivotal role: already in Paper I  we have shown that at least the average rotational velocity of the stars must be limited to $<50$ km/s at metallicities [Fe/H]$>-1$, in order to avoid an overproduction of  heavy nuclei, mainly in the Ba peak. But there are also other subtle indirect factors that may change the yields predicted by rotating models: in order to bring protons in an He active environment, at least part of the H rich mantle must be present while He is burning. A substantial change in the mass loss rate (e.g. due to the inclusion of a dust driven component to the mass loss rate or to the overcome of the Eddington luminosity) may affect the range of masses that retain a substantial fraction of the H rich mantle while the stars are in the central He-burning phase. 

Uncertainties of stellar models is one of the reasons  why the validity of the stellar method has been questioned  \citep[see e.g.][]{arn07}. Another one is that this method does not consider the solar s-(r-)process abundance distribution in an astrophysical framework, i.e. as the result of all the previous generations of stars which polluted the interstellar medium  prior to  the formation of the Solar system. In particular, these generations of stars covered a large range of metallicities and {\it not a unique value (or even a limited range of values) of [Fe/H]} as it is assumed in the classical and stellar methods. For instance, it is well known that at low metallicities a large neutron/seed ratio is obtained, leading to the production of the heaviest s-nuclei, while at high metallicities the opposite happens \citep[see e.g.][and references therein]{Tra04}. 

The Solar system s-(r-) process abundances have to be understood  in the framework  of a galactic chemical evolution (GCE) model. This is certainly a difficult task that requires a good understanding of the star formation history in the Galaxy, of stellar evolution, and of the interplay between stars and the interstellar gas, among other things. We are still far from fully understanding these issues. Therefore, this third method is based on a necessarily schematic description of the situation considering the chemical evolution of our Galaxy, accounting for the fact that the site(s) of the r-process have not been clearly identified yet. 
 Attempts to obtain the s- and r- components of the solar composition from a GCE model were pioneered by \citet{Tra04}, later  updated by \citet{ser09} and more recently by \citet{Bi14,Bi17}. These authors employed a GCE code adopting s-process yields
 from AGB stellar models by \citet{Gal98} in a range of masses and metallicities (see these papers for details). Regarding the r-process yields, and for elements from Ba to Pb, they estimated the contribution to the Solar system by subtracting the s-residuals from the solar abundances. Then, they scale the r-process yields to the yield of a primary element (in a similar way we do here, see Eq. 4) mainly produced in core collapse supernovae, which they assumed to occur in the mass range $8-10$ M$_\odot$. They derived the weak s-process contribution from \cite{Rai93}. On the other hand, for the lighter elements, in particular for Sr-Y-Zr, they deduced the r-residuals and, thus, the r-process yields, from the abundance pattern found in  CS 22892-052 (\citealt{sne02}),
 by assuming that
 the abundance signatures of this star is of {\it pure r-process origin} (i.e., any contamination by other possible stellar sources is hidden by the r-process abundances). 
 
 The \citet{Bi14} model resulted in good agreement with the Solar s-only isotopic abundances between $^{134,136}$Ba and $^{204}$Pb, also showing that the solar abundance of $^{208}$Pb is well reproduced by metal-poor AGB stars, without requiring the existence of a "strong" component in the s-process as is done in the classical method. Below the magic number $N=82$, however, they found a significant discrepancy between the abundance distribution obtained with their GCE model and the Solar system values.  It turned out that their GCE model underproduces the solar s-process component of the abundances of Sr, Y and Zr by $\sim20\%-30\%$ and also the s-only isotopes from $^{96}$Mo up to $^{130}$Xe. This result prompted \citet{Tra04} to postulate the existence of another source of neutron-capture nucleosynthesis named the {\it light element primary process (LEPP)}. They argued that this  process is different from the s-process in AGB stars and also different from the weak s-process component occurring in massive stars. The recent updates of this study by \citet{Bi14,Bi17}, reach the same conclusion\footnote{\citet{Bi14,Bi17} mainly focus on the impact of the different $^{13}$C pocket choices in AGB stars and weak s-process yields from massive stars, on the s-process residuals at the epoch of the Solar System formation. In \citet{Bi17} yields from massive stars are included considering the impact of rotation in a limited range of masses and metallicities according to the models by \cite{Fri16}.}. In particular, these two studies ascribe a fraction ranging from $8\%$ to $18\%$ of the solar Sr, Y and Zr abundances to this LEPP, and suggest that lighter elements from Cu to Kr could be also affected. On the other hand, they obtained a r-process fraction at the Solar system ranging from $8\%$ (Y) to $50\%$ (Ru). 
 
 The need of a LEPP has been recently questioned by \citet{Cr15b} and later by \citet{Tr16} on the basis of a simple GCE model using updated s-process yields from AGB stars
\citep{Cr11} and AGB stellar models only, respectively. These studies show that a fraction of the order of that ascribed to the LEPP in the predicted solar abundances of Sr, Y and Zr can be easily obtained, for instance, by just a moderate change in the star formation rate prescription in a GCE model, still fulfilling the main observational constrains in the solar neighbourhood. The same effect can be found by modifying AGB stellar yields as due to nuclear uncertainties, or the choice of the mass and profile of the $^{13}$C pocket. Introducing such a changes in the GCE models (i.e. stellar yields) one can easily account for the {\it missing} fractions of the solar abundance of  these elements within the observational uncertainties.
In addition, in Paper I we have very recently shown that the LEPP is not necessary  when metallicity-dependent s-process yields from rotating massive stars (i.e. the "weak" s-process)  are considered in a GCE model. The  stellar yields adopted in that paper are from an extended grid of stellar masses, metallicities and rotation velocities from \cite{lim18}\footnote{As stated in Paper I and \cite{lim18}, the nuclear network for massive stars includes 335 isotopes in total, from H to$^{209}$Bi, and is suited to properly follow all the stable and explosive nuclear burning stages of massive stars. The portion of the network from H to $^{98}$Mo takes into account all the possible links among the various nuclear species due to weak and strong interactions. For  heavier nuclei, we consider only (n,$\gamma$) and $\beta$-decays. Since we are mainly interested in following in detail the flux of neutrons through all the magic number bottlenecks and since in the neutron capture chain the slowest reactions are the ones involving magic nuclei, between $^{98}$Mo and $^{209}$Bi we explicitly follow and include in the nuclear network, all the stable and unstable isotopes around the magic numbers corresponding to N=82 and N=126 and assume all the other intermediate isotopes at local equilibrium.}; for the first time in GCE studies, the IDROV was introduced through an empirically determined function of metallicity and rotation velocity. 

In this study, we use the GCE model of 
Paper I to derive the s- and r-process contributions to the solar isotopic abundances in the full  mass range from $^{69}$Ga to $^{235}$U through a new method.

\section{The method}
\label{sec:OurMethod}

\subsection{s-only and r-only isotopes}
\label{subsec:S-R-only}

The classification of nuclei belonging to the {\it s-only} group is not a trivial task.  By definition, an s-only nucleus owes its entire abundance to the slow neutron capture process. As a consequence,
we tentatively identify as s-only any nucleus with atomic number Z for which a stable isobar with atomic 
number Z-1 (or Z-2) exists: that isobar  shields the nucleus from any r-process contribution. However, such a condition is necessary, but not sufficient to define 
an s-only isotope. In fact, there are isotopes lying on the proton-rich side of the $\beta$-stability valley, that are shielded
from the r-process, but may receive a non negligible contribution from the p-process 
\citep[see][]{tra15}.
Moreover, there are isotopes with unstable isobars with (Z-1), whose lifetimes are comparable to the
age of the Universe: in that case, therefore, a delayed r-process contribution cannot be excluded (e.g. for the couples $^{87}$Sr-$^{87}$Rb 
and $^{187}$Os-$^{187}$Re). Finally, there are a few isotopes, with stable (Z-1) isobars, which may receive an important contribution from the neutrino process in core collapse supernovae (e.g. $^{113}$In and $^{115}$Sn; see
\citealt{fuj07}).
As a matter of fact, in the past different lists of s-only isotopes circulated in the literature. We list in Table \ref{tab:s-only} the s-only isotopes considered in this study, including those that may receive a small contribution from other processes \citep[see][]{tra15}.

 The definition of r-only isotopes in even more ambiguous. In principle, at odds with s-only nuclei (shielded by the r-process from stable isobars), there is no nucleus fully shielded by the s-process. In fact, all nuclei on the neutron-rich side of the $\beta$-stability valley can receive a contribution (perhaps very small, but not null) from the s-process, depending on the activation of various branchings. For instance, net yields from AGB stars by \citet{Cr15} for isotopes marked as r-only in previous compilations (e.g. \citealt{gor99} and \citealt{sne08}) are all positive (from some \% to significant fractions, depending on the isotope), apart from $^{130}$Te. In this study we shall not pre-define "r-only" nuclei, but we shall explore with our method the contribution of our stellar yields to the abundances of all heavy isotopes.

\begin{table}
	\centering
	\caption{List of 30 s-only isotopes adopted in this work}
	\label{tab:s-only}
	\begin{tabular}{ccl} 
		\hline
		Z & Isotope & Possible contribution  \\
		\hline
32 & $^{70}$Ge & \\
34 & $^{76}$Se & \\
36 & $^{80}$Kr & from p-process \\
36 & $^{82}$Kr & \\
38 & $^{86}$Sr & from p-process  \\
38 & $^{87}$Sr &  from $^{87}$Rb\\
42 & $^{96}$Mo & \\
44 & $^{100}$Ru & \\
46 & $^{104}$Pd & \\
48 & $^{110}$Cd & \\
50 & $^{116}$Sn & \\
52 & $^{122}$Te & \\
52 & $^{123}$Te & \\
52 & $^{124}$Te & \\
54 & $^{128}$Xe &  \\
54 & $^{130}$Xe &\\
56 & $^{134}$Ba & \\
56 & $^{136}$Ba &\\
60 & $^{142}$Nd & \\
62 & $^{148}$Sm & \\
62 & $^{150}$Sm & \\
64 & $^{154}$Gd & \\
66 & $^{160}$Dy &  \\
70 & $^{170}$Yb &\\
71 & $^{176}$Lu & radiogenic  to $^{176}$Hf \\
72 & $^{176}$Hf & radiogenic  from $^{176}$Lu \\
76 & $^{186}$Os & \\
78 & $^{192}$Pt &\\
80 & $^{198}$Hg & \\
82 & $^{204}$Pb & \\
		
\hline
	\end{tabular}
\end{table}

\subsection{Assumptions}
\label{subsec:Assumptions}

The method adopted in this study is based on a couple of key assumptions.

{\bf Assumption 1}: Our current understanding of stellar nucleosynthesis and galactic chemical evolution allows us to reproduce the pre-solar isotopic abundances to a precision of (a) a factor of $\sim 2$ for elements  with charge 2 $<$ Z $<$ 30 (between Li and Zn) but (b) to {\it a factor of $\sim$20-30 \% (or less) for the s-component of heavier elements}.

Statement (a) above is based on the fact that all calculations done up to now with "state-of-the-art" stellar yields and models of the chemical evolution of the solar neighborhood show indeed a dispersion of a factor $\sim$2  around the solar value  in the region up to the Fe-peak. This is true e.g. for the models of \cite{Timmes1995}, who adopted yields of \cite{Woo95}, \cite{Goswami2000} with yields of \cite{Woo95}, \cite{Kubryk2015a} with yields of \cite{Nomoto2013} and Paper I with yields of \cite{lim18}. Even if in each case the adopted models and yields differ considerably, the outcome is the same: a dispersion by a factor of $\sim$2 is always found, implying that uncertainties in the various parameters of the problem (regarding both stellar and galactic physics) remain important in the past two decades or so.

Statement (b) is based on a limited sample of GCE models, namely
those of \cite{Tra04}, \citet{Cr15b} and \cite{Bi17} - see previous section -  as well as our own model presented in Paper I. In those
by \cite{Tra04} and \cite{Bi17}, the model values of most heavy pure s-nuclei barely exceeds the corresponding solar value and there is a systematic deficiency of $\sim$20-30\%  as one moves to lighter s-nuclei. This deficiency was interpreted  as  evidence for the need of another heavy isotope component, the so-called LEPP (see previous section). However,  Paper I showed that rotating massive stars may produce through the weak s-process that "missing" component, with no need for a new process. In that study, it is found that most pure s-nuclei are co-produced within $\sim$10-20\% from their pres-solar values, with only a few of them displaying higher values (up to 40\% at most).

We think that it is illusory at the present stage of our knowledge to reproduce the pre-solar pure s-composition to a higher accuracy. We believe however that it is possible to use this result and try to infer the solar s- and r-components of all mixed (s+r) nuclei, as presented in Sec. \ref{subsec:Method}.   

{\bf Assumption 2}: {\it The r-process is of "primary" nature and, in particular, it mimics the behaviour of the "alpha" process which produces $\alpha$-elements like e.g. $^{16}$O. } This assumption is based on the observational fact that pure r-elements, like Eu, display an $\alpha$-like behaviour, i.e. the ratio [Eu/Fe] remains $\sim$constant at a value of $\sim$0.3-0.5 dex during the late halo evolution and then declines smoothly to its solar value at [Fe/H]$\sim $0. This means that, in contrast to the s-process, which is basically of "secondary" nature (i.e. the s-yields of both LIM stars and massive stars depend on the abundance of iron-seed nuclei), the r-yields are independent of the initial metallicity of their source. The ratio of those yields to the yields of $\alpha$-isotopes should then be constant with metallicity. These inferences allow one to adopt r-process yields "scaled" to the stellar model yields used in the GCE model.

\subsection{Method}
\label{subsec:Method}

\begin{figure*}
	\includegraphics[width=0.75\textwidth, angle=-90]{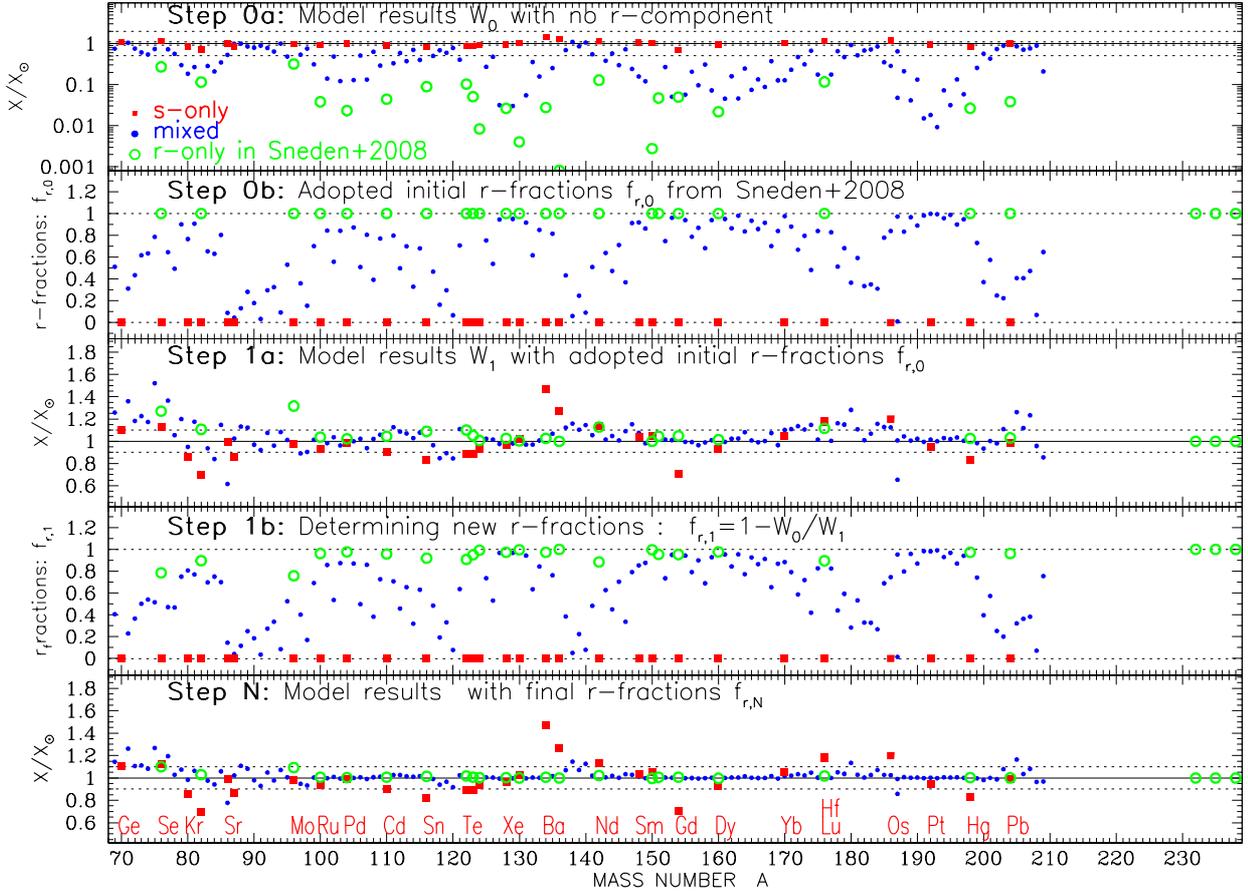} 
	\caption{ {\it Top:} Model results $W_0$ after Step 0, without r-component (see text); horizontal dashed lines indicate levels of $\pm$10\% and a factor of 2 deviation from solar;   
	2$^{\rm nd}$ from top: Adopted initial r-fractions from  \citet{sne08};
	3$^{\rm rd}$ {\it from top}: Results $W_1$ after Step 1, with r-component introduced from \citet{sne08} ; horizontal dashed lines indicate levels of $\pm$10\%  deviation from solar; 4$^{\rm th}$ {\it from top}: Our r-component after Step 1 is obtained as $r_1 = 1-W_0/W_1$ (where the corresponding s-component is obtained first as $s_1=W_0/W_1$) and is introduced in the next iteration; {\it Bottom:} Same as the 3d panel, after the final (N=17 here) iteration of our "bootstrap" method; horizontal dashed lines indicate levels of $\pm$10\%  deviation from solar. Note that the scale in the Y-axis changes in the different panels.The  names of the elements with s- only isotopes are indicated in the bottom panel.}
\label{fig:Method} 
\end{figure*}

Our "bootstrap" method proceeds as follows:

{\bf {\it Step 0}}: We run a GCE model as in Paper I but  using exclusively the s-component for all elements with Z$>$30. For that purpose we remove from the adopted yields 
the r-component (i.e. existing in the initial composition of the stars, through their scaled solar composition). In practice, we calculate a new set of yields as:
\begin{equation}
y_i (M,Z) \ = \ y_{i,0}(M,Z) \ -  \ f_{r,i,0} \ X_{i,\odot} \ Z \ M_{ej}(M,Z)  
\end{equation}
where:
\begin{itemize}
    \item $y_{i,0} (M,Z)$ are the original yields of isotope $i$ from  stars of mass $M$ and metallicity $Z$.
    \item $f_{r,i,0}$ is the solar r-fraction of nucleus  $ i $, as provided i.e. in \cite{sne08} or \cite{gor99}.
    \item $ M_{ej}(M,Z)$ is the total mass ejected by the star of mass $M$ and metallicity $Z$.
\end{itemize}
The results of the model at the time of Solar system formation (i.e. 4.56 Gyr before the end of the simulation) are stored as 
\begin{equation}
  W_{i,0} \ = \ X_{i,0}/X_{i,\odot}
\label{eq:Step0}
\end{equation}
i.e., they are normalized to the corresponding Solar system isotopic abundances adopted from \citet{Lod09}. These normalized abundances appear in the top panel of Fig. \ref{fig:Method} for all nuclei with charge Z$>$30. 

Among the s-only nuclei (red dots), most are reproduced within a factor of 20\% solar abundances (see also Paper I \footnote{Notice that with respect to Paper I we have slightly reduced here the proportion of fast rotating massive stars (at 300 kms$^{-1}$) in our mixture, in order to avoid an overproduction of the lighter s-only nuclei like $^{70}$Ge and $^{76}$Se; this reduction affects correspondingly the results of $^{80,82}$Kr (compare e.g. to Fig. 11 in Paper I) but no other nuclei, either lighter or heavier ones), since they are essentially produced by LIM stars.}), except Kr, Ba and Gd which differ from their solar values by 20-40 \%.Taking into account the uncertainties in nuclear, stellar and galactic physics involved in the calculation, which lead to a larger dispersion for the lighter nuclei (up to 100 \% , factor of $\sim$2, see Fig. 11 in Paper I), we think that this agreement is quite satisfactory. 

In particular, regarding the nuclear uncertainties, we note that our results are obtained with nucleosynthesis calculations using the set of neutron capture cross sections described in  \citet{str06}. Since then, two new cross sections became available, i.e. those of $^{176}$Lu \citep{wis062} and $^{176}$Hf \citep{wis06}. Both cross sections are larger than those adopted to calculate our models, so that we expect a decrease for both isotopes (see Table \ref{tab:s-only}), thus providing a better agreement with observations. We expect a similar behavior for $^{134}$Ba (and possibly $^{136}$Ba): both neutron capture cross sections will be measured in the next years at the n\_TOF facility \citep{gue13}. Moreover, we further stress that the abundance of $^{134}$Ba strongly depends on the activation of the branching at $^{134}$Cs, whose neutron capture cross section and temperature-dependent $\beta$-decay lifetime are rather uncertain. By varying theoretical nuclear inputs within uncertainties in a single model, we can obtain a decrease of about 15\% and 12\% for $^{134}$Ba and $^{136}$Ba, respectively \citep[see also][]{Cr15b,gor99}. All the above concern s-only isotopes which are over-produced with respect to their pre-solar system values in Fig. \ref{fig:Method}.

As for the s-only isotopes that are under-produced with respect to the solar distribution, we stress that the neutron capture cross section of $^{82}$Kr is quite uncertain at typical s-process temperatures ($\sim$25\% at 8 keV; KADONIS database\footnote{https://exp-astro.de/kadonis1.0/}).  Moreover, it has to be stressed that the solar Kr and Xe abundances are not directly measured in the Sun, but they {\it" are based on theoretical values from neutron-capture element systematics"} \citep{lo03}. On the other hand, the synthesis of $^{154}$Gd is strongly affected by the branching at $^{154}$Eu. Its neutron capture cross section has never been measured and its temperature-dependent $\beta$-decay lifetime is uncertain by a factor of three \citep{gor99}. Note that for the decay, no hints on its trend between 5$\times 10^7$ K and laboratory temperature is provided in \cite{tak87}. As already done for barium isotopes, if we just vary theoretical nuclear inputs within uncertainties, we can obtain an increase of about 25\% for $^{154}$Gd\footnote{Note that the $^{154}$Gd neutron capture cross section has been recently measured at the n\_TOF facility (Massimi et al., in preparation)}.

Finally, it should be emphasized  that we did not make any attempt to adjust the parameters of the GCE model (distribution of stellar rotational velocities, initial mass function or star formation and infall rates) as to optimize the s-only distribution; as discussed in Paper I, our GCE model is tuned in order to reproduce as well as possible local parameters like the current gas fraction, the metallicity distribution and age-metallicity relation and the abundances of major elements like O and Fe at Solar system formation. Despite that, we find that the parameter  
\begin{equation}
 \rm    g = exp~\left[ {1\over n_{S}}  \sum_{Z,A}^{n_{S}} ln^2 {N_{cal}(Z,A)\over N_\odot(Z,A)}\right]^{1/2}
\end{equation}
where the sum runs over the $n_{S}=30$ s-only nuclei (of charge Z and mass A, see Table 1) is $g=1.18$, i.e. it is not much higher than the value of 1.10 obtained in  \cite{gor99}. This author optimized the few parameters of his multi-event model as to minimize $g$, while we did not attempt such an optimization here. (with a classical analysis \cite{gor99} found  $g=1.44$). 
Although our results are obtained with a different method and data (nuclear cross sections, stellar conditions, solar abundances), we believe that our result regarding the s-only distribution is quite reasonable and constitutes a good starting point for our GCE method. We discuss a little more our distribution of the s-component of our GCE model in Sec. 4, where we present the resulting $\sigma_A$N$_A$ distribution.

We emphasize here that, in contrast to  the GCE method of \cite{Tra04,Bi14,Bi17}  we do not proceed directly after the first run to the evaluation of the solar s-component by subtracting our results from the solar composition. This might lead to the need of a LEPP to justify the underproduction of several s-only nuclei, as the aforementioned GCE studies did. We proceed in a different way, allowing us to keep the "s-only" property of the isotopes of Table 1 and at the same time evaluate self-consistently the s-fraction of the mixed (s+r) isotopes. For that, we {\it need} to introduce  {\it a priori}  their
r-fractions, as  described below.

{\it Step 1:} We run a model by using now the original stellar yields $y_{i,0}(M,Z)$ and introducing this time  the r-component of each isotope as in Paper I, namely by assuming that it is co-produced with a typical product of massive stars like $^{16}$O, i.e. the new yield {\it for massive stars } (M$>$10 \ms) is
\begin{equation}
  y_{i,1}(M,Z) \ = \ y_{i,0}(M,Z) \ + \ f_{r,i,0} \ y_{\rm ^{16}O}(M,Z) \ X_{i,\odot}/X_{\rm ^{16}O,\odot}  
\label{eq:Ryields}
\end{equation}
where the last term represents the r- component of the yield and $f_{r,i,0}$  is an "educated guess" for the solar r- fraction of isotope $i$; we start by adopting the r- fractions of  \cite{sne08} but our results are independent of that choise (see below). 

The underlying physical assumption of Eq. \ref{eq:Ryields} is that $^{16}$O and the r-component have the same source, namely massive stars and this implicit assumption allows one to reproduce naturally the observed alpha-like behaviour of elements that are mostly of r-origin, like e.g. Eu.
The method can be used in essentially the same way in the case that the main source of r-process turns out to be a rare class of massive stars, like collapsars \citep[see e.g.][and references therein]{siegel2019}. In that case a stochastic treatment should be made, e.g. as applied for neutron star mergers in   \citet{Ojima2018}.
If neutron star mergers are assumed to be the site of the r-process, a different prescription should be used, involving the rate of occurrence of that site (through a delayed time distribution, as for SNIa, e.g. \citealt{cot18}) and  the mass ejected  in the form of isotope $i$, normalized as to get a solar abundance for the pure r-isotopes of Th and U.

Notice that in this run we treat {\it all} nuclei except the s-only ones of Table 1 as mixed s+r: those classified as pure r- in \cite{sne08}   or \cite{gor99} are also treated as such. They are simply given an {\it initial r-fraction} $f_{r,i,0}$=1, which may change after Step 1.
    
The result of the new run is also plotted in Fig. \ref{fig:Method} for all isotopes with charge Z$>30$ as overabundances 
\begin{equation}
  W_{i,1} \ = \ X_{i,1}/X_{i,\odot}
\label{eq:Overab}
\end{equation}  
where
\begin{equation}  
 X_{i,1} \  = \ X_{s,i,1} \ +  \ X_{r,i,1} 
\label{eq:tot-comp}
\end{equation}
with $X_{s,i,1} = X_{s,i,0}$ (the s-component remains the same) and 
\begin{equation}  
X_{r,i,1}/X_{i,\odot}  = C \ f_{r,i,0} 
\label{eq:r-comp}
\end{equation}
is the r-component (proportional to the r-fraction $f_{r,i})  $ with the constant $C$ being the  IMF average of the r-component term in Eq. \ref{eq:Ryields} and
 adjusted as to obtain at Solar system formation the exact solar abundances of pure r-isotopes, like Th,   which we use here as benchmarks\footnote{The radioactive decay of Th and U isotopes is properly taken into account in our GCE model.}. The value of  {\it C} depends on the adopted ingredients of the GCE model (IMF, SF and infall rates, stellar yields) and it is $\sim$1.12 in our case. 
 
 One notices that:
\begin{itemize}
    \item s-only isotopes are produced exactly at the same level as in step 0, since their yields have not been modified.
    \item r-only isotopes  with the meaning discussed in \S \ref{subsec:S-R-only} are produced exactly at their pre-solar abundances -because of the adopted normalization in Eq. \ref{eq:Ryields} and \ref{eq:r-comp} - except a few of them which  have received
a non-zero contribution from the s-process in step 0 (see green symbols in top panel of Fig. \ref{fig:Method}) and are now slightly overproduced. The most prominent of them are $^{76}$Ge (by $\sim$25\%), $^{82}$Se ($\sim$10\%), $^{96}$Zr  ($\sim$30\%) and $^{142}$Ce ($\sim$15\%), as shown in the $3^{rd}$ panel of Fig. \ref{fig:Method}.
    \item isotopes of mixed (s+r) origin are nicely co-produced w.r.t. their pre-solar abundances, to better than $10\%$ in general, although in some regions (A$\sim 205, 180, 138, <95$) they are overproduced by $\sim$20\% and the overproduction reaches 40\% for the lightest ones.
\end{itemize}

Obviously, by comparing the results of runs 0 and 1 (top    and  third from top panels) one may obtain the s-fraction of each mixed isotope as
\begin{equation}
f_{s,i,1} \ = \ W_{i,0}/W_{i,1} \ = \  W_{i,0}/(W_{i,0} \ + \ C  \ f_{r,i,0})
\label{eq:s_fraction}
\end{equation}
and the corresponding r-fraction as
\begin{equation}
 f_{r,i,1} \ = \ 1 - f_{s,i,1} \ = \ 1 - \ W_{i,0}/(W_{i,0} \ + \ C  \ f_{r,i,0})
 \label{eq:r_fraction}
\end{equation}

This procedure was adopted in Paper I, albeit not for the pure r-isotopes for which {\it we assumed} a final r-fraction equal to the initial one $f_{r}$=1. However, at this level the  method was obviously not self-consistent: the resulting r-residuals, obtained with Eq. \ref{eq:r_fraction} were not the same as those used to run the model with the r-component in Eq. \ref{eq:Ryields}. This is obvious in the 4$^{th}$ panel of Fig. \ref{fig:Method}, in particular regarding the r-fractions of $^{76}$Ge (which is now $\sim$80\% instead of 100\% initially) and $^{82}$Se (now $\sim$90\% instead of 100\%).

{\it Step 3}: In this study, seeking for self-consistency, we proceed by injecting the obtained r-fractions of  step 1 and Eq. \ref{eq:r_fraction} into the yields of Eq. \ref{eq:Ryields} and running a new model. The results of the new model are identical with those of previous calculations regarding all isotopes below Z$=30$ and  the pure s-ones, but they fit slightly better the pre-solar distribution of mixed s+r isotopes. 

We evaluate the quality of the fit to the solar composition through a simple $\chi^2$ test and we repeat running the models injecting each time the new r-fraction obtained through Eq. \ref{eq:r_fraction} into the yields of mixed isotopes. The fit improves slower and slower as the number of iterations increases, until the improvement becomes negligible (less than 1 part in a thousand) and we stop. This happens in general  after 10-20 iterations, depending on the initial r-fractions adopted\footnote{The number of iterations required to reach a given level of convergence increases with decreasing $f_{r,0}$; for a level of 10$^{-2}$ we find that 20-30 iterations are sufficient.}.

From the mathematical point of view, it can be easily shown that the quantities $W$ (Eq. \ref{eq:Overab}),  $f_s$ (Eq. \ref{eq:s_fraction}),  and $f_r$ (Eq. \ref{eq:r_fraction}), expressed as recursive functions,  converge  to the values indicated in Table \ref{tab:limits1}, depending on whether the constant $C$ is greater or smaller than the initial overabundance $W_0$\footnote{This can be trivially obtained by putting $f_r$=$f_{r,0}$ in Eq. \ref{eq:r_fraction}}, i.e. the s- component.  In other terms, our results for the s- and r- fractions depend uniquely on a) the adopted stellar yields of s-isotopes (which determine, along with the chemical evolution model, the term $W_0$), and b) the goodness of the fit to the pure r- isotopes of Th and U (which determine through Eq. \ref{eq:Ryields} and \ref{eq:r-comp} the constant $C$), but they are independent of the choice of the initial values of $f_{r,0}$. 

The reason why this iterative method improves - albeit slightly - the overall fit is due to the fact that the sum of the s- and r- fractions for a mixed nucleus is always $f_s + f_r$=1. If the new s-fraction is found (Eq. \ref{eq:s_fraction}) to be smaller than the original one, then the new r-fraction is automatically found to be larger than the original one to compensate, and vice versa.

In the bottom panel of Fig. 1 we display the results of the final run. The agreement with pre-solar abundances is now considerably improved for the mixed (s+r) nuclei, which are reproduced to better than a few \% in most cases. We consider this a satisfactory result and we believe that it is the best one may hope to get from current models of stellar nucleosynthesis and galactic chemical evolution.

We also repeated the procedure by adopting the initial r-residuals of \cite{gor99} and we obtained quantitatively similar results for all mixed (s+r) isotopes, except for the few cases which are classified as pure s- or r- by \cite{gor99} but not by \cite{sne08}; these are cases where the minor residual has a very small contribution to the isotopic abundance, typically less than a few \%, which may be smaller than the uncertainties defined by the method of \cite{gor99}.  

\begin{table}
	\centering
	\caption{Limits of recursive functions $W, f_s$ and $f_r$ for $k \rightarrow \infty$}
	\label{tab:limits1}
	\begin{tabular}{lcc} 
		\hline
		Function  & $C > W_0$ & $C \leq W_0$  \\
		\hline
 $W_k$ \ = \ $W_0 \  + \ C \ f_{r,k-1}$ & $C$  &  $W_0$ \\
 
 $f_{s,k}$ \ = \ $\frac{W_0}{W_0 \ + \ C  \ f_{r,k-1}}$ & $\frac{W_0}{C}$    & 1 \\
 
 $f_{r,k}$ \ = \ 1 \ - \ $\frac{W_0}{W_0 \ + \ C  f_{r,k-1}}$  & 1 \ - \ $\frac{W_0}{C}$ & 0 \\
\hline
\end{tabular}
\end{table}

\section{Results and discussion}
\label{sec:Results}

Our results concerning the s- and r- fractions of all the heavy isotopes are presented in Table \ref{tab:long}, along with those of \cite{gor99} and \cite{sne08} as well as those of \cite{Bi14}; notice that for the latter we provide only the s-contribution (see below). For an easier comparison with those studies, the data are also presented in Figs. \ref{fig:Only s-r} and  \ref{fig:f_comparison}. Although it is impossible (and rather meaningless) to perform a one-to-one comparison for each isotope, we notice some important features.

\subsection{The s- and r- fractions}
\label{subsec:S-fractions}

\begin{figure}
\begin{centering}
\includegraphics[width=0.49\textwidth]{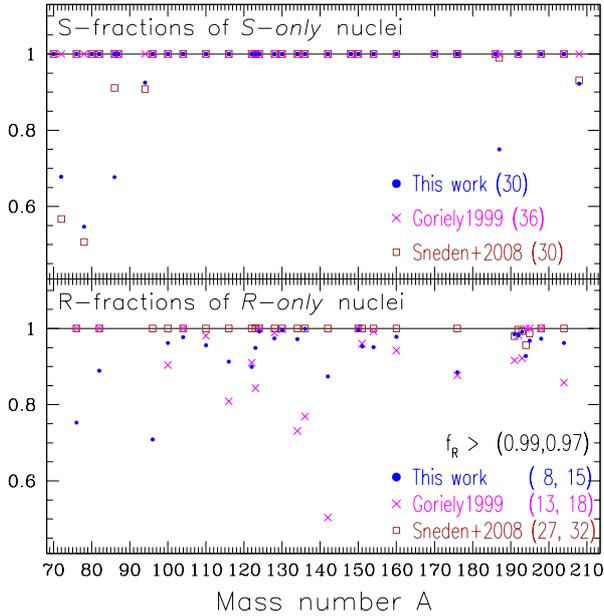}

\caption{ {\it Top:} Nuclei having an s-fraction $f_s$=1  in at least one of the lists of  \citet{gor99}, \citet{sne08} or ours. {\it Bottom:} Nuclei with  r-fraction $f_r=1$ in at least one of the cited studies. In parentheses: the number of such nuclei in each study. In the bottom panel: two numbers are given, for r- fractions f$_r >0.99$ and $>0.97$, respectively; the Th and U isotopes are counted in, even if they do not appear in the figure.}
\label{fig:Only s-r}
\end{centering}
\end{figure}

We start by displaying in Fig. \ref{fig:Only s-r} the results for heavy nuclei that have been classified as s-only (top) or r-only (bottom) in each of the studies of \cite{gor99}, \cite{sne08} and the present one. We emphasize that in our study the nuclei considered as 
s-only in the beginning (Model 0) are also found to be s-only during the whole procedure and in the final model, since the adopted yields for those species are always the same (exactly as in the case of nuclei lighter than  Z$=31$). This does not mean that their final abundances match perfectly well the corresponding Solar system abundances. But we consider that the obtained deviations from the solar abundances are a natural feature of the adopted GCE method, reflecting the current limitations of 1-zone models of GCE (coming mainly from stellar yields).

\begin{figure}
\begin{centering}
\includegraphics[width=0.49\textwidth]{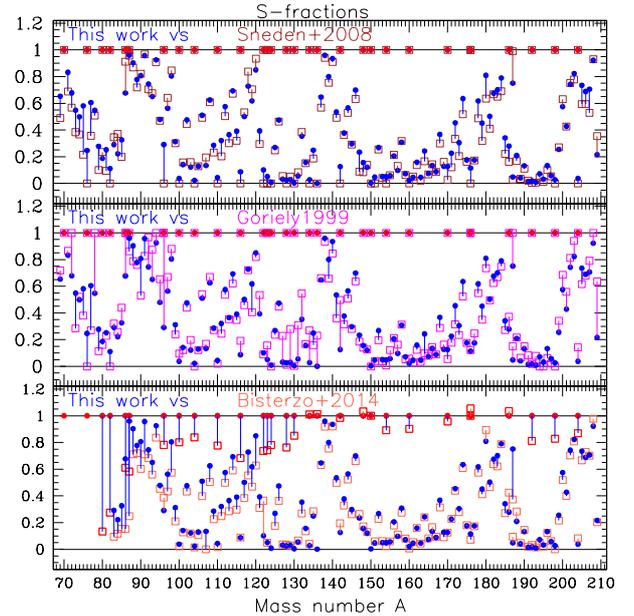} 
\par\end{centering}
\caption{\label{fig:f_comparison} Our s-fractions (in blue in all panels) compared to results of three  works obtained with different methods. {\it Top:} \citet{sne08} with the classical method. {\it Middle:} \citet{gor99} with the multi-event method. {\it Bottom:} \citet{Bi14} with a GCE model.
Vertical lines connect same nuclei, their colour corresponding to the largest s-fraction of the two results. Red circles indicate our s-only nuclei (see Table  \ref{tab:s-only}).}
\end{figure}

Our method (dividing M0 by M1) allows us to attribute $f_s$=1 to those nuclei that we {\it pre-defined} as s- only. This is not the case with the other studies using GCE models \citep{Tra04,Bi14}, which try to reproduce perfectly the solar abundances of s-only isotopes, something we think is illusory at present. This is why we postpone the discussion of our differences with those studies to Fig. \ref{fig:f_comparison}. 

We  notice that the s- and r-fractions displayed in Fig. \ref{fig:Only s-r} are evaluated in different ways for the three studies. \citet{sne08} provide the absolute numbers for both components s- and r- ($\rm N_s$ and $\rm N_r$) for each heavy isotope, in a scale where $\rm N_{Si} \equiv$10$^6$; in that case one has obviously: $\rm f_s$=$\rm N_s$/($\rm N_s$+$\rm N_r$) and $\rm f_r$=$\rm N_r$/($\rm N_s$+$\rm N_r$). \citet{gor99} provides only the $\rm N_r$ values, again in a scale $\rm N_{Si} \equiv$10$^6$, but he does not provide the corresponding $\rm N_s$ values; we obtain here the corresponding  $\rm N_s$ values by subtracting $\rm N_r$ from the total isotopic abundances $\rm N_T$ where we use the pre-solar ones of \cite{Lod09}, which were not available in 1999. This obviously introduces some systematic differences with the actual values found by \citet{gor99}, hopefully small ones.   We do not take into account a couple of nuclei considered as s-only in  \cite{gor99}, which we consider instead as p-nuclei, like $^{152}$Gd and $^{164}$Er.

Fig. \ref{fig:Only s-r} illustrates the difficulties to determine unambiguously whether an isotope is produced exclusively by one  or the other of the two neutron capture processes. While \citet{gor99} finds 36 s-only isotopes, we and \citet{sne08} find only 30. For two of the six discrepant cases, $^{94}$Zr and $^{208}$Pb, we and \citet{sne08} find quite high s-fractions of more than 90\%, i.e. almost pure s-nuclei. For two others ($^{72}$Ge and $^{78}$Se) we both find a dominant s-contribution of 55-65 \%, which leaves room however, for a large r-contribution. Finally, there are two cases
($^{86}$Kr and $^{187}$Os) where  \citet{sne08} find a very high s-contribution of more than 90\% (making those nuclei almost s-only), while we find a smaller one, around 70-75\%. It is hard to trace the exact origin of these differences, which can be broadly attributed to the different methods and data followed to derive the s- fractions. The abundance of $^{86}$Kr is determined by the branching at $^{85}$Kr;  its activation largely depends on the temperature and, as a consequence, requires the use of full stellar models to be properly treated. On the other hand, $^{187}$Os may receive an important contribution from the decay of $^{187}$Re during the long burning phases of low and intermediate mass stars (the so-called "astration" term of \citealt{yokoi83}). Unfortunately, such a decay is not known at temperatures intermediate between laboratory values and some tens of millions K, making the evaluation of the "astration" term in full stellar evolutionary models very uncertain.

As for the r-only nuclei, the bottom panel of Fig. \ref{fig:Only s-r} shows that their number varies strongly with the adopted criterion for their definition. If an 
r-contribution f$_r>0.99$ is required, then only 8 nuclei fulfill it in our case, against 13 for  \citet{gor99} and 27 for \cite{sne08}. If f$_r>0.97$ is adopted instead, the corresponding numbers become 15, 18 and 32, respectively. Obviously, this reflects directly the difficulty to determine accurately  the corresponding s-fractions, which are fairly small. The reason for the discrepancy between \cite{sne08} and the other two studies is quite probably the broader range of physical conditions (temperature and neutron fluence) spanned in the multi-event study and this work, which allows the neutron flow to reach in some cases nuclei usually unreachable in the "classical" method. The most characteristic cases are $^{134,136}$Xe for which \citet{gor99} finds an r-contribution of $\sim$75\% (against more than 97\% in our case), and $^{142}$Ce where he obtains 50\% only while we obtain 87\% (see Table \ref{tab:long}). 

\begin{figure*}
\begin{centering}
\includegraphics[angle=-90,width=0.9\textwidth, viewport=253 5 593 760,clip=true]{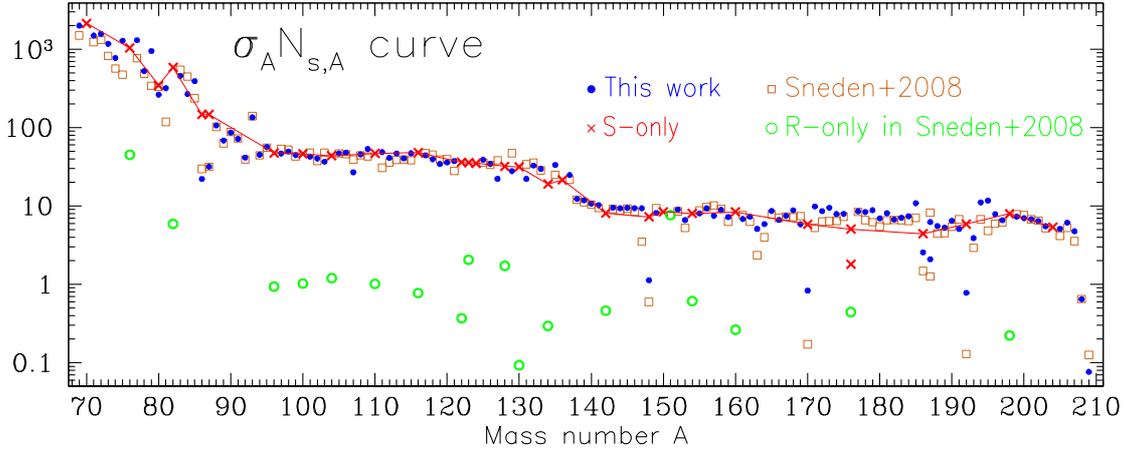}

\caption{\label{fig:f_sigmaN} Curve $\sigma _A$N$_{s,A}$ of the pre-solar system s-component, after  \citet{sne08} ({\it brown open squares}) and this work ({\it all other symbols}). Our s-only nuclei appear as {\it red  crosses} and are connected by a solid curve to guide the eye. The  {\it green open circles}  represent  nuclei that are 
r-only for \citet{sne08} (hence they should not appear on the figure), but they receive a small s- contribution in our work.  Cross-sections are expressed in mb and number abundances ${\rm N}_{s,A}$ are in the meteoritic scale of ${\rm N}_{\rm Si} \equiv 10^6$.}
\end{centering}
\end{figure*}

The most significant discrepancy between those studies is the case of $^{96}$Zr. It is considered as a pure r-nucleus in \citet{sne08}, but  we find only 70\% while \citet{gor99} finds that its 
r-contribution is compatible with zero, i.e. that it may be  a pure s-nucleus. The abundance of this nucleus strongly depends on the treatment of the branching at $^{95}$Zr, which is activated during thermal pulses only (i.e. when the temperature at the base of the convective shell exceeds $2.5\times 10^8$ K). We notice that, in our models, the contribution of the s-process (from LIM stars) is 30\%, which is compatible with the 40 \% s-contribution of \cite{Bi14}. 

In Fig. \ref{fig:f_comparison} we present a more detailed comparison of our results to those of \citet{sne08} (top panel), \citet{gor99} (middle) and \cite{Bi14} (bottom). We notice that there are  few s-only nuclei (i.e. with f$_s\sim$1) in the latter work, since they are not "pre-defined", as in our case. Thus, despite the  use of a similar method to ours (GCE), the 
s-component of the heavy nuclei is identified in a different way in \cite{Bi14} and this makes more difficult a direct comparison to our results. It turns out, however, that apart form the 
s-only nuclei, our results display largely similar features.

In the range A$<85$, our results are systematically higher than \citet{sne08} or \cite{Bi14} and rather  closer to \citet{gor99}. This is  due to the fact that our adopted yields from rotating massive stars (responsible for the weak s-process in that region of mass number) produce abundantly the light s-nuclei: as explained in Paper I, rotational mixing of $^{14}$N from the H-layer in the He-core produces more $^{22}$Ne than in non-rotating models and enhances substantially the neutron fluency and the resulting  s-isotope production (see also \citealt{cho16,cho18}). This leads to a larger s-process contribution to the isotopic abundances in that mass range.

 The cases of $^{76}$Ge and $^{82}$Se illustrate well theses findings. They are classified as pure r-nuclei by \citet{sne08} 
and \citet{gor99} 
whereas our method leads to a 25\% s-contribution to the former and a 11 \% s-contribution to the latter, because of the rotating massive star yields.

In the region $85<$A$<200$ our results are in fairly good quantitative agreement with \citet{sne08}, while \citet{gor99} finds systematically higher s-fractions for $85<$A$<100$ and $120<$A$<135$.  However, in all these cases the uncertainties in the determination of the s-fractions (as evaluated only by \citealt{gor99}) are substantial and the results can be considered as compatible with each other. This is also true for several cases found to be pure r-isotopes by \citet{gor99}, while both \citet{sne08} and us find a small s-contribution (e.g. $^{153}$Eu,  $^{159}$Tb, $^{161}$Dy).

In the region A$>200$ \citet{gor99} has, in general, larger s-contributions than both \citet{sne08} and us. His multi-event method results in a stronger "strong" s-process component than the other studies.

Above A$=100$ and up to the heaviest nuclei, our results are in excellent agreement with \cite{Bi14}, except for the cases of the s-only nuclei (already discussed). This is the case of  $^{142}$Ce which is a pure r-nucleus for \cite{sne08}, while we find an s-contribution of $\sim$13\% and \cite{Bi14} find a 20\% 
s-contribution. Finally,  $^{187}$Os has a much larger 
s-contribution of 75\% in our case vs 37\% in \cite{Bi14}, while it is a pure s-nucleus in both  \cite{sne08} and \cite{gor99}.

\subsection{The $\sigma _A \rm N_A$ curve}
\label{subsec:sigmaN}

The classical method to determinate the s-component of the heavy isotopes relies on the assumed constancy of the product  $\sigma_A \rm N_{s,A}$ of the neutron capture cross section $\sigma_A$ times the number abundance N$_{s,A}$ of the 
s-component of the heavy nuclei.  In Fig. \ref{fig:f_sigmaN} we display that product for our results, obtained from our s-fraction from Table \ref{tab:long}, the corresponding Solar system isotopic abundance from \cite{Lod09} (also provided in Table \ref{tab:long}) and the neutron capture cross sections at 30 keV provided in the KADONIS database\footnote{Online at http://www.kadonis.org} \citep{Dillmann2008}.

Our $\sigma_A \rm N_{s,A}$ curve displays the "classical" features, namely, a decrease up to A$\sim$90, a near constant value up to the 2$^{nd}$ peak at A$\sim$135 (the Ba isotopes), then a small decline and again an approximately constant value up to the 3$^{rd}$ peak at A$\sim$205 (the Pb isotopes). We notice, however, that the $\sigma_A \rm N_{s,A}$ product of 
s-only nuclei (red cross symbols, connected with a solid curve) is not constant in the whole range of A before the 3$^{rd}$ peak, but it declines by almost a factor of 2 between A$=160$ and A$=190$;  this would make it difficult to derive accurately with the classical method the s-only abundances of isotopes in that mass range, at least with the  sets of $\sigma_A $ and $\rm N_{s,A}$ adopted here.

\begin{figure*}
\begin{centering}
	\includegraphics[angle=270,width=0.9\textwidth,viewport=173 5 593 760,clip=true]{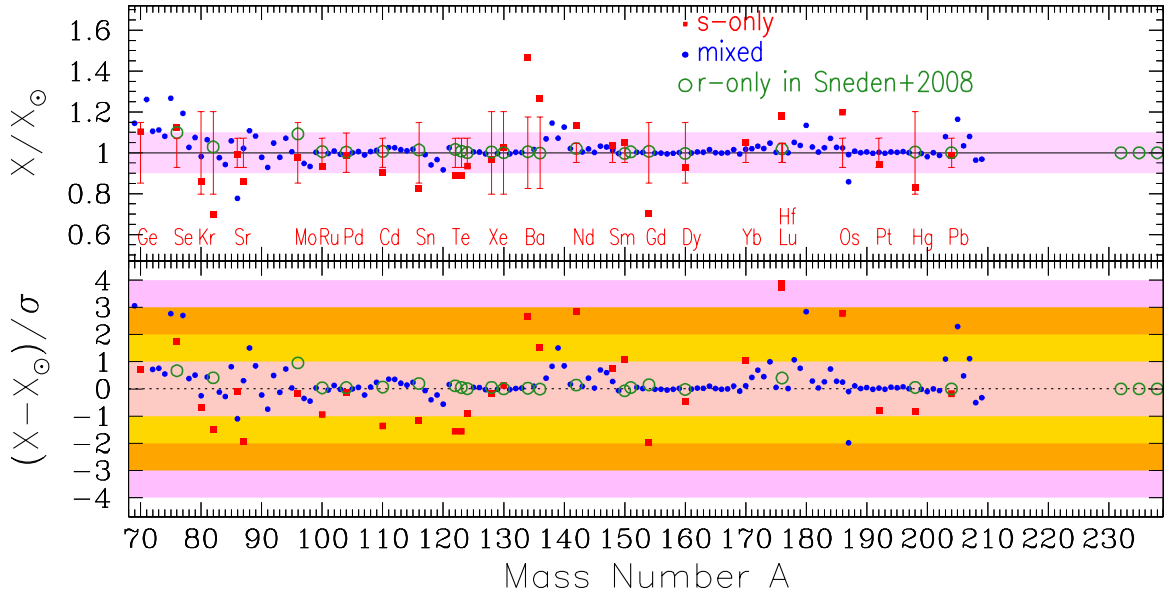}
\par\end{centering}
\caption{\label{fig:f_uncertainties} Results compared to uncertainties of pre-solar  abundances. {\it Top:} Model over-abundances at the time of Solar system formation and elementary uncertainties $\pm \rm 1 \sigma$ for the s- only isotopes (vertical red segments, centered at X/X$_{_\odot}$=1 from \citet{Lod09}, the  names of which are provided in the bottom of the panel; the shaded area indicates the range of $\pm$10\%  deviation from solar. {\it Bottom:} Overabundances (positive) or underabundances (negative) expressed in terms of the corresponding elementary uncertainties. Shadowed areas indicate ranges of $\pm \rm 1\sigma$, $\pm \rm 2\sigma$, etc.}
\end{figure*}

 In Fig. \ref{fig:f_sigmaN} we also present  the $\sigma_A \rm N_{s,A}$ product obtained with the s-component of \citet{sne08},  multiplied by the same set of cross-sections as in our case. In that way, the differences between the two sets of results depend only on the s-residuals, which are obtained by two different methods. Although the s-residuals of \citet{sne08} were obtained {\it by using the $\sigma_A \rm N_A$ method   with a different set of cross-sections} (older than the one used here and presumably less accurate), still it is interesting to compare the two sets of results.

There is fairly good agreement for a large range of mass numbers ($90 <$ A $< 190$). There are, however, important discrepancies (by  factor of $\sim$2) in the range of light s-nuclei, for $74<$A$<87$; this is not surprising, since  $\sigma_A \rm N_{s,A}$ is not expected to be constant in that atomic mass  region, making it difficult to derive the s-component with the classical method.

The most important differences between the two results concern:
\begin{itemize}
    \item The three nuclei $^{76}$Ge, $^{82}$Se and  $^{96}$Zr, which are pure r-nuclei (open circles in the figure) in both \citet{gor99} and 
    \citet{sne08} but are found to receive an s-contribution of 25\%, 11\% and 30\%, respectively, from our s-process, the first two from rotating massive stars and the third from LIM stars.
    \item Several other nuclei, which are considered pure r-nuclei in \cite{sne08} but have s-contribution in our case and in \cite{gor99}. In  most cases that contribution is of a few \%, but it mounts to 10\% for $^{116}$Cd, $^{142}$Ce and $^{176}$Yb; similar results for the s-contribution of all those nuclei are obtained in \cite{Bi14}. 
    \item Nuclei lying near branching points, like $^{148}$Nd, $^{170}$Er and $^{192}$Os, which receive a fairly small 
    s-contribution in \citet{sne08} but a considerably larger one (factors 2-4 larger) in our case 
 \end{itemize}

    We notice that in the last two cases, our results agree fairly well with those of \citet{gor99}  and \cite{Bi14}, probably because these studies explore more realistic physical conditions in stellar interiors than the classical study of \citet{sne08} could do. We consider this is an important advantage of  those methods over the classical one in determining the  s-fractions - and, thereoff, the 
    r-fractions - of the solar composition.

\subsection{Comparison to Solar system abundances}
\label{subsec:Comparison_SS}

In the previous sub-sections we presented our s- and r-fractions of the heavy isotopes and we compared them to those of previous studies, pointing out similarities and discrepancies. In some cases we were able to attribute those discrepancies in differences in the adopted data and methods. We also presented our  $\sigma_A \rm N_{s,A}$ curve, which constitutes an important criterion of the validity of our method; we showed that it succeeds fairly well and in some cases (e.g. the nuclei in branching points) even better than the classical method. 

In this sub-section we present the
abundances of all heavy isotopes (already displayed in the bottom panel of Fig. \ref{fig:Method}) and we compare them to the Solar system ones, taking into account the measurement uncertainties of the latter. As we  emphasized in Paper I, our results strongly depend on the adopted yields of LIM stars and rotating massive stars, but also on the adopted chemical evolution model, because of the extreme sensitivity of the s-process yields to stellar metallicity.  We also reiterate here two of our main findings in Paper I, namely that a) rotating massive stars contribute the bulk of the weak s-process up to A$\sim$90  (and very little above it) while low mass stars are major s-contributors for A>90, and b) we find no compelling evidence for the LEPP invoked in \cite{Tra04}, especially if uncertainties in measured Solar system abundances are taken into account (see Fig. 11 and Sec. 3.2.2 of that paper).

In the top panel of Fig. \ref{fig:f_uncertainties} it is seen that the vast majority of the heavy isotopes (120 out of a list of 149, from which the p-isotopes are excluded) lie within $\pm$10\% of their solar values, while 96 of them lie within $\pm$5\%. However, among the 30 s-only nuclei, only half (14) lie within $\pm$10\% of their solar values and a quarter (8) lie within $\pm$5\%, while 4 display a deviation (overabundance or underabundance) between 10-20\%, 3  show a deviation of 20\% to 30\% and one ($^{134}$Ba) a deviation of 45\%.

We notice that the dispersion of the pure s-nuclei is substantially larger than the dispersion of the other heavy nuclei, as expected:  the latter have an adjustable component (the r-component), while the former are the direct product of the adopted ingredients for stellar nucleosynthesis and chemical evolution of the solar neighborhood. We also notice that the deviations of the  distribution of the s-nuclei from the solar one drives the deviations of the mixed (s+r) nuclei: the latter are important in the "weak" s-process region (A$<90$), between $^{116}$Sn and the  $^{122,123,124}$Te isotopes, just above the  $^{132,134}$Ba isotopes and in the $172<$A$<182$ region. The "strong" s-component, at A$\sim 205$ also appears enhanced.

The dispersion of the heavy nuclei is clearly smaller than the one of the isotopes lighter than the Fe-peak, as described in Paper I. The reason is that the latter are produced in various advanced phases of massive star evolution (including very poorly understood ones as the final stellar explosion) while the s-nuclei are produced in the better understood phase of He-burning.  On top of that, the introduction of an "adjustable" (through our iteration procedure) r- component further improves the situation for the heavies.

The true magnitude of  the deviation of the model from the observed solar composition is also understood in terms of the uncertainties in the measured solar abundances. We notice here that measured uncertainties in \cite{Lod09} concern elemental abundances, not isotopic ones, and we assume here that they apply to all the isotopes of a given element. In the top panel of Fig. \ref{fig:f_uncertainties} these uncertainties are displayed as vertical error bars for the s-only nuclei. The bottom panel of the same figure displays the situation for all the heavy nuclei in a different way: over- (or under-) abundances are presented in units of the corresponding 1$\sigma$ uncertainty. Now 120 isotopes are found within $\pm$1$\sigma$, 18 within  $\pm$2 $\sigma$ and 7 within $\pm$3$\sigma$ of their solar values, whether the 2 isotopes of A$=176$ (Lu and Hf) are at almost 4$\sigma$.

\begin{figure*}
\begin{centering}
	\includegraphics[angle=270,width=0.95\textwidth,viewport=173 5 593 760,clip=true]{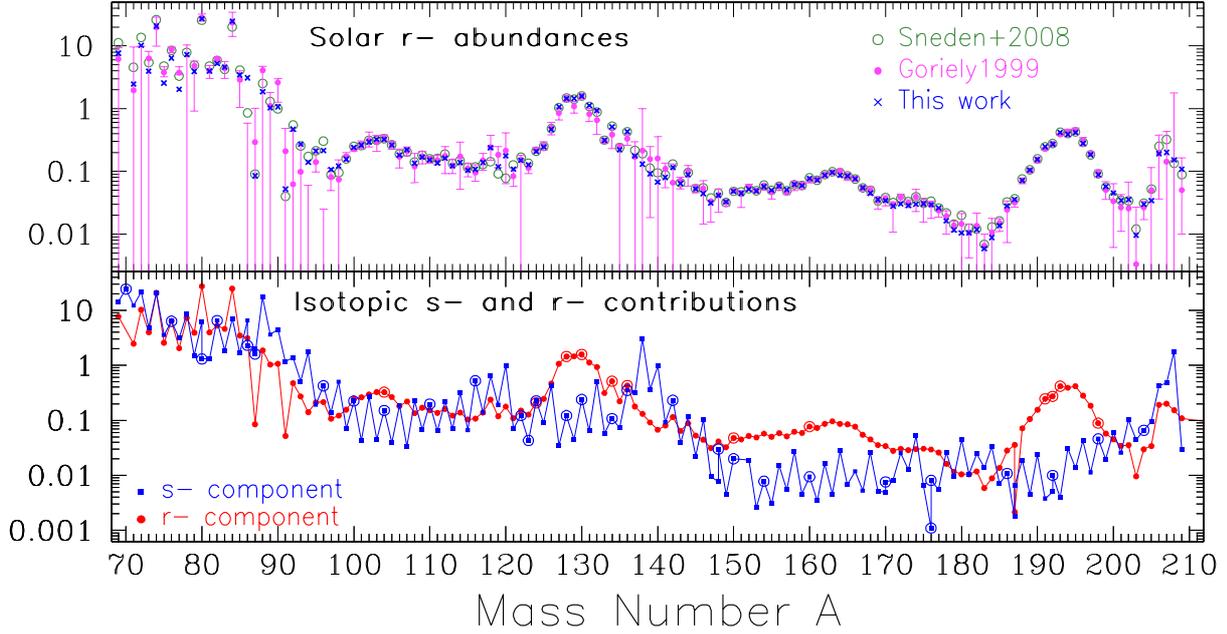}
\par\end{centering}
    \caption{{\it Top:} Absolute r-residuals, i.e. contribution by number of r-process N$_{r}$ to pre-solar composition  compared to \citet{sne08} and \citet{gor99}. Number abundances are expressed in the meteoritic scale of N$_{\rm Si}\equiv$ 10$^6$. {\it Bottom}: s- and r-contributions to Solar system isotopic composition according to this work. Open circles represent s-only or r-only nuclei (see text).}
    \label{fig:r-component}
\end{figure*}

All values within $\pm$1$\sigma$ of the observed ones, i.e. the vast majority of heavy nuclei, can be considered as perfectly reproduced by the model, to the present level of our knowledge. The majority of the remaining ones (16 out of 28) are s-only nuclei. As already discussed, it seems difficult  to further reduce the dispersion of those nuclei around their solar values in the framework of present-day stellar nucleosynthesis and GCE models. Further developments (e.g. concerning nuclear inputs,  improved treatment of the $^{13}$C pocket in LIM stars or of the rotational mixing in massive stars, a better understanding of the contribution of rotating stars to the chemical evolution of the Galaxy, etc.) will certainly help to reduce that dispersion. The method presented here will allow us then to determine completely and accurately the contribution of the s- and 
r-processes to the solar composition in a realistic global astrophysical framework.

\subsection{The r- component}
\label{subsec:R-component}

Our r-residuals, i.e. the isotopic Solar system abundances of \cite{Lod09} multiplied with the r-fractions derived in this work (as presented in Table \ref{tab:long}) are displayed in Fig. \ref{fig:r-component}. They are compared (top panel) to
those derived by \citet{gor99} and \citet{sne08}. The former study includes the model uncertainties, resulting from the corresponding  uncertainties in the observed Solar system composition, the neutron radiative capture rates (n,$\gamma$) and the $\beta$-decay rates. The uncertainties in those quantities date back to more than 20 years ago and some of them have been reduced in the meantime. However, the work of \cite{gor99} is the only one up to now to include a systematic evaluation of those uncertainties and we chose to display them here in order to provide some idea of their importance.

A first glance at the top panel of Fig. \ref{fig:r-component} (and a quantitative one in Table \ref{tab:long}) shows that there is a fairly good overall agreement between the three studies for the region $100 < $A$ < 200$: a simple $\rm \chi^2$ test for the N$=85$ isotopes (excluding the s- only) gives $\rm \chi^2/$N$\sim$0.03 when comparing our results to both \citet{gor99} and \citet{sne08}. This is also the region with the smallest number of uncertain r-residuals in the study of \citet{gor99}.

In the regions A$<100$ (N$=30$) and $200<$A$<210$ (N$=9$) we obtain clearly better agreement with \citet{sne08} ($\rm \chi^2$/N$\sim0.06$ for A$<100$ and $\rm \chi^2$/N$\sim0.02$ for A$<200$) than with \citet{gor99} ($\rm \chi^2$/N$\sim0.30$ for A$<100$ and $\rm \chi^2$/N$\sim0.22$ for A$<200$).  Our discrepancies are produced essentially for nuclei with fairly small r-fractions, like $^{88}$Sr, $^{91,94}$Zr (where we obtain r-fractions of a few percent whereas \citet{gor99} obtains $\sim$20\% on average). In all those cases, however, the uncertainties quoted in \citet{gor99} are quite large, making our results compatible with his. For  $^{75}$As and $^{77}$Se, our r-residuals  are lower than in both \citet{gor99} and \citet{sne08}, lying below the quoted uncertainties of the former. This is due to our enhanced s-component from rotating massive stars, leading to a low r-component for those nuclei.  As already discussed,  $^{86}$Kr and $^{96}$Zr are cases apart, since \citet{gor99} finds a zero r-contribution while all other studies in Table \ref{tab:long} find a substantial r-fraction, lying even outside his quoted uncertainties.

We did not attempt here any evaluation of the uncertainties of our model results, but in the previous section we compared them to the solar abundances and took into account the corresponding uncertainties in the abundance measurements. It is interesting to notice here the broad similarity between   some features in Figs. \ref{fig:f_comparison} and \ref{fig:r-component} (top panel). The regions of the largest deviation of our results from solar in Fig. \ref{fig:f_comparison} are for A$<100$, A$\sim$140 and A$\sim$205; those same regions display the largest uncertainties in their r-component in \citet{gor99}, which in turn are driven by the uncertainties in the corresponding s-fractions. More work on these A regions is required to further reduce those uncertainties.

\begin{figure*}
\begin{centering}
	\includegraphics[angle=270,width=0.95\textwidth,viewport=173 5 593 760,clip=true]{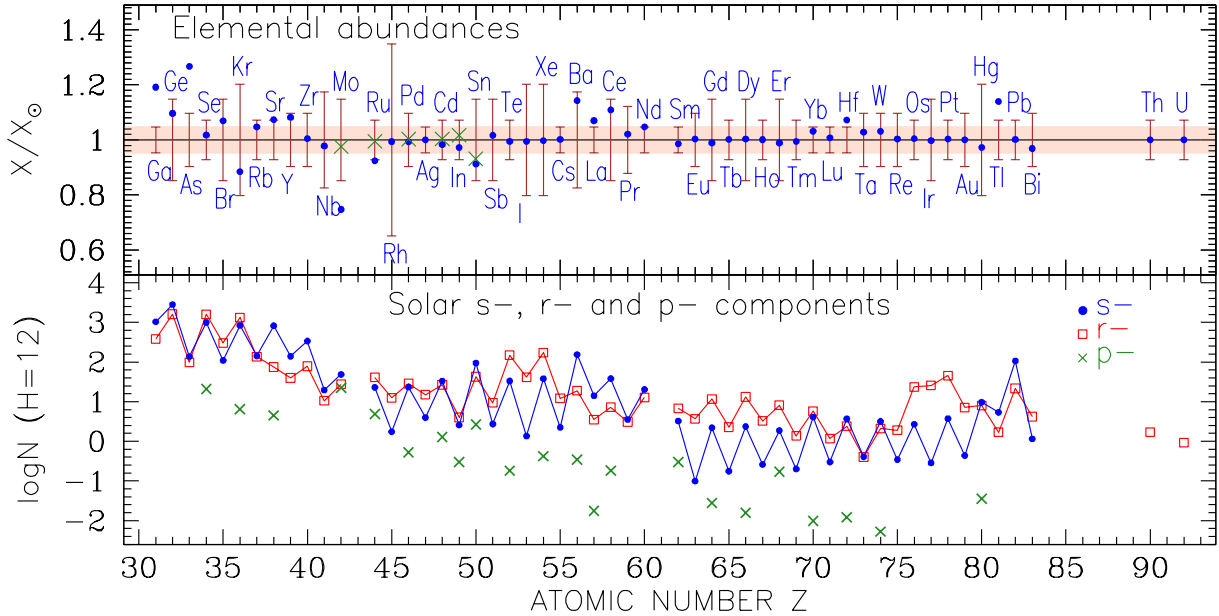}
\end{centering}	
    \caption{{\it Top:} Elemental pre-solar abundances for Z>30 obtained with the final GCE model at the Solar System formation 4.5 Gyr ago and compared to the data of \citet{Lod09} (vertical uncertainty bars).The green crosses correspond to the addition of the p-component (23\% in the case of Mo, see Table \ref{tab:Elements}). The horizontal shaded area indicates the range of $\pm$5\% around the solar values.  
    {\it Bottom:} Solar abundance pattern by number  decomposed into s-(blue points), r-(red squares) and 
    p-(green crosses) components according to this study. }
\label{fig:Elms}    
\end{figure*}

Finally, in the bottom panel of Fig. \ref{fig:r-component}  we display the isotopic s- and r-contributions to the Solar system composition according to our results. The distributions differ little from previous ones - especially in the adopted logarithmic scale. Pure r-nuclei are designated with red open circles in that panel and correspond to r-fractions f$_r>0.97$. As discussed in \S \ref{sec:Results} their definition remains ambiguous, and we adopt here a rather arbitrary criterion accounting for the estimated uncertainties.

\subsection{Elemental  abundances }
\label{subsec:Elemental_anundances}

In Fig. \ref{fig:Elms} we plot the elementary abundances of our model. In the top panel, we compare them to the solar abundances of \cite{Lod09}. 36 out of the 51 elements from Ga to Bi lie within $\pm$5\% of their Solar system values (we do not count Th and U, they are reproduced exactly at their solar value by construction, but their role is important because they are used to monitor the way we introduce the r-process isotopes in our GCE calculations).

\begin{figure*}
\begin{centering}
	\includegraphics[angle=270,width=0.9\textwidth]{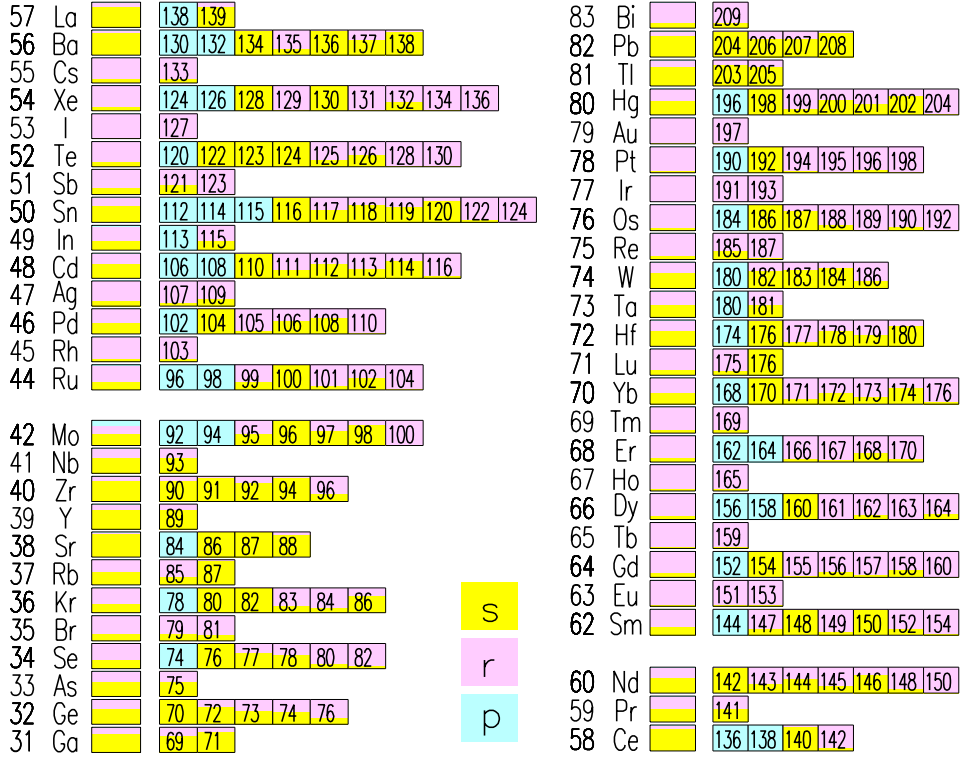}
	\end{centering}
    \caption{Contribution of the s-, r- and p-process to the isotopic and elemental pre-solar composition according to   
      this work (from Tables \ref{tab:long} and \ref{tab:Elements}). 
    The contribution of each process is proportional to the colored area of the corresponding box.}
    \label{fig:final}
\end{figure*}

The above result is obtained without considering the role of the p-isotopes. The 35 p-isotopes are a minor component of the Solar system composition. They are shielded from the neutron capture processes\footnote{The s-process may contribute to the production of the p- nuclei $^{152}$Gd and $^{164}$Er, as well as to $^{115}$Sn, $^{180}$Ta and $^{180}$W; see \citet{Arnould2003} and references therein.} and they are supposed to be produced by either the p- (proton captures) or the $\gamma$- (photo-desintegration of heavier nuclei) processes; the former may play some role in the production of a few of the lighter p-nuclei while the latter dominates for most p-isotopes  \citep{Arnould2003}. Despite decades of theoretical and experimental studies, their origin, nucleosynthesis sites and galactic chemical evolution are still poorly understood (see \citealt{Travaglio2018}  and references therein). We did not include those isotopes in our study, but we have to include them when comparing our final elementary composition with the Solar system abundances. 

For that purpose we consider all the 35 p-isotopes as unaffected by any neutron capture process and we simply add their contribution (from \citealt{Lod09}, as appearing in our Table \ref{tab:Elements}) to our s+r contributions of each element at Solar system formation. The results to the handful of elements affected by that addition appear as crosses in the top panel of Fig. \ref{fig:Elms}. Mo receives an important contribution (23\%)
 from its $^{92,94}$Mo isotopes and rutenium a small one (7\%) from $^{96,98}$Ru. Pd, Cd, In and Sn also receive contributions of 2-4\% from their p-isotopes.
 The addition of those  contribution from p-isotopes improves largely the situation for Mo and Ru, which are now also produced to better than 3\% of their Solar system abundance. 
 
 When the uncertainties in the experimental determination of the Solar system composition are taken into account, the situation improves largely. Only 3 out of the 51 elements are clearly above 1 $\sigma$ of their solar abundance (Ga, As and Tl), while La and Hf are barely above 1 $\sigma$. The conclusion is that  our understanding of the solar composition in heavy elements is fairly satisfactory, except for the lighter of those elements which lie in the region of the weak s-process. Rotating massive stars appear to be important contributors to that region, but their role has still to be explored and better constrained. The more so since other sources, poorly explored up to now, may also contribute in that mass region, like electron-capture supernovae \citep[e.g.][]{Wanajo2011},  the {\it $\nu $p}-process \citep[e.g.][]{fro06} and the {\it rp}-process \citep[e.g.][]{sch98}.

 Our results do not reproduce perfectly the solar composition, but we assume that they reflect our current understanding of the situation. The interest of our method lies in the fact that it allows us to evaluate self-consistently the fractional contribution of the s- and r-processes to the solar abundances (while the p-fraction is directly taken from observations). Even if some of the s-only isotopes are overproduced in our calculation, the adopted method leads automatically to an s-fraction equal to one for them, and it evaluates similarly the s-fractions (and then the
 r-fractions) of the mixed nuclei.
 In the bottom panel of Fig. \ref{fig:Elms} we display then our decomposition of the solar system elemental abundance distribution into s-, r- and p-components, multiplying the corresponding fractions from Table \ref{tab:Elements} with the solar elemental abundances of \cite{Lod09}. 

 Fig. \ref{fig:final} reproduces the ensemble of our results concerning the elemental and isotopic contributions of the three nucleosynthetic processes to the Solar system composition in an original way, allowing one to grasp at a glance the importance of one of the three nucleosynthetic processes to the corresponding abundance. 

\section{Summary}
\label{sec:Summary}

In this work we present a new method for assessing the s- and r-fractions of the Solar system abundances of  heavy nuclei (heavier than Zn). 

Our method (Sec. \ref{subsec:Method}) is based on a GCE model using state-of-the-art yields of LIM stars and rotating massive stars, for all isotopes from H to Bi. The model has been presented in detail in Paper I and, as extensively discussed there, satisfies all the main observational constraints that can be expected from a 1-zone model  for the solar neighborhood. In particular, the isotopic distribution of nuclei up to the Fe-peak at Solar system formation (4.5 Gyr ago) is well reproduced.

Our method consists in running first a model with only the s- component for all heavy isotopes, i.e.  without considering any r-component in the stellar ejecta (model M0). Then, the r-component is introduced (model M1) based on some prior estimate of it (in our case, the r-fractions estimated by \citealt{gor99} or \citealt{sne08}). We assume that the r-process is primary in nature and follows the evolution of an alpha element, as suggested by observations of e.g. Eu. The s-fractions of all nuclei $f_s$ are thus obtained as the results of M0/M1 at Solar system formation. They are by construction $\leq$1 (independently of whether the corresponding isotopes are overproduced w.r.t. their solar abundance) and they are equal to one for the 30 heavy isotopes which are {\it pre-defined} as s-only (Table \ref{tab:s-only}). This is one of the advantages of our GCE method. The corresponding r-fractions are defined as $f_r$=1-$f_s$, they are also  positive (independently of whether the corresponding nuclei in Model 0 or 1 are overproduced or not) and they differ slightly from their initial r-fraction (the adopted prior).

In order to make our GCE model  self-consistent, we then inject the new
r-fractions into it and run it again. The new isotopic composition fits better the solar one (as indicated by a simple $\chi^2$ test) and we iterate again with the new r-component until the results do not vary sensitively any more. We obtain thus a final decomposition of the Solar system isotopic abundance distribution into s- and r-components, fully self-consistent with our GCE model and the adopted stellar yields.

We present our results in Table \ref{tab:long} and we compare them to those of previous studies, based on different methods \citep{gor99,sne08,Bi14} in Sec. \ref{subsec:S-R-only}. 
We find good overall agreement between the various studies, but also some  discrepancies. The most important are found in the region of the "weak s-process", namely below A$=90$. In that region, \citet{sne08} find that $^{76}$Ge and $^{82}$Se are r-only nuclei (in agreement with \citealt{gor99}), while we find a substantial contribution from the s-process in rotating massive stars (25\% and 11\%, respectively).

Our $\sigma$N$_{s,A}$ curve (first time derived from a GCE model) displays  the classical feature of $\sigma$N$_{s,A}\sim$const between magic neutron numbers, but shows an interesting difference with the classical study of \citet{sne08} (Sec. \ref{subsec:sigmaN}):
nuclei lying near branching points, like $^{148}$Nd, $^{170}$Er and $^{192}$Os, receive a fairly small  s-contribution in \citet{sne08} but a considerably larger one (factors 2-4) in our case;  our results are, in general, in better agreement with those of \citet{gor99} and \cite{Bi14}, probably because their studies explore a larger range of (and/or more realistic) physical conditions in stellar interiors than the classical study of \citet{sne08}.

Comparison of our model distribution with measured solar abundances (\S \ref{subsec:Comparison_SS}) shows an excellent agreement, especially when uncertainties in the measured abundances are considered. The most important deviations concern the s-only nuclei, because there is no possibility to modify the result by an adjustable r-component, as we do for all other mixed (s+r or pure r-) nuclei. Since the final abundances are "driven" by the s-component of Model 0, it is clear that the most important deviations are expected in the regions where the s-process dominates, namely the three s-peaks (see Fig. \ref{fig:f_comparison}). This is encouraging, since it implies that a better treatment of the s-process in both LIM stars and rotating massive stars will allow one to reduce further those deviations from the solar composition by applying our method. We notice here that the study of \citet{gor99} finds that the largest uncertainties in the
r-component (displayed in Fig. \ref{fig:r-component}, top panel) concern precisely those regions. In \S \ref{subsec:Elemental_anundances} we present our results for the s-, r- and p- components of heavy elements in the Sun, both in graphical (Fig. \ref{fig:Elms} and \ref{fig:final}) and tabular (Table \ref{tab:Elements}) forms. 


In summary, we propose a new method for evaluating the s- and r-components of the chemical composition of the Sun and during the Milky Way history, in a way fully consistent with our current understanding of stellar nucleosynthesis and galactic chemical evolution. The accuracy of our results obviously depends on the current uncertainties in those fields, as well as on the uncertainties in the measured solar isotopic composition.

\section*{Acknowledgements} This article is based upon work partially supported from the “ChETEC” COST Action (CA16117) of COST (European Cooperation in Science and Technology). C.A. acknowledges in part to the Spanish grants AYA2015-63588-P and PGC2018-095317-B-C21 within the European Founds for Regional Development (FEDER).

\clearpage
\onecolumn

\begin{small}

\setlength{\tabcolsep}{2pt}

\begin{longtable} {c c c c c c c c c c }

\caption{Contribution (by mass fraction) to the Solar system isotopic composition by the s- and r-process, as obtained in \citet{sne08}, \citet{gor99}, \citet{Bi14} (only for the s-process) and this work. Solar system abundances N$_{\odot}$ from \citet{Lod09} are given per $10^6$ Si atoms.} 
\label{tab:long}
\\
\hline
      \multicolumn{3}{c}{Isotope} &
      \multicolumn{2}{c}{Sne+2008} &
      \multicolumn{2}{c}{Gor1999} &
       \multicolumn{1}{c}{Bis2014} &      
      \multicolumn{2}{c}{This Work} \\
Z & A & N$_{\odot}$ & s- & r- & s- & r- & s- & s- & r-  \\
\hline
\endfirsthead
\hline
\multicolumn{10}{|c|}{Continuation of Table \ref{tab:long}}\\
\hline
      \multicolumn{3}{c}{Isotope} &
      \multicolumn{2}{c}{Sne+2008} &
      \multicolumn{2}{c}{Gor1999} &
      \multicolumn{1}{c}{Bis2014} &   
      \multicolumn{2}{c}{This Work} \\
Z & A & N$_{\odot}$ & s- & r- & s- & r- & s- & s- & r-  \\
\hline
\endhead
  31 &$^{ 69}$Ga & 2.20E+01 & 0.490 & 0.510 & 0.719 & 0.281 & - & 0.653 & 0.347 \\
  31 &$^{ 71}$Ga & 1.46E+01 & 0.689 & 0.311 & 0.866 & 0.134 & - & 0.832 & 0.168 \\
  32 &$^{ 70}$Ge & 2.43E+01 & 1.000 & 0.000 & 1.000 & 0.000 & - & 1.000 & 0.000 \\
  32 &$^{ 72}$Ge & 3.17E+01 & 0.567 & 0.433 & 1.000 & 0.000 & - & 0.678 & 0.322 \\
  32 &$^{ 73}$Ge & 8.80E+00 & 0.384 & 0.616 & 0.283 & 0.717 & - & 0.548 & 0.452 \\
  32 &$^{ 74}$Ge & 4.12E+01 & 0.366 & 0.634 & 0.522 & 0.478 & - & 0.499 & 0.501 \\
  32 &$^{ 76}$Ge & 8.50E+00 & 0.000 & 1.000 & 0.000 & 1.000 & - & 0.247 & 0.753 \\
  33 &$^{ 75}$As & 6.10E+00 & 0.215 & 0.785 & 0.380 & 0.620 & - & 0.581 & 0.419 \\
  34 &$^{ 76}$Se & 6.32E+00 & 1.000 & 0.000 & 1.000 & 0.000 & - & 1.000 & 0.000 \\
  34 &$^{ 77}$Se & 5.15E+00 & 0.356 & 0.644 & 0.270 & 0.730 & - & 0.605 & 0.395 \\
  34 &$^{ 78}$Se & 1.60E+01 & 0.507 & 0.493 & 1.000 & 0.000 & - & 0.547 & 0.453 \\
  34 &$^{ 80}$Se & 3.35E+01 & 0.235 & 0.765 & 0.161 & 0.839 & - & 0.187 & 0.813 \\
  34 &$^{ 82}$Se & 5.89E+00 & 0.000 & 1.000 & 0.000 & 1.000 & - & 0.111 & 0.889 \\
  35 &$^{ 79}$Br & 5.43E+00 & 0.100 & 0.900 & 0.114 & 0.886 & - & 0.278 & 0.722 \\
  35 &$^{ 81}$Br & 5.28E+00 & 0.094 & 0.906 & 0.229 & 0.771 & - & 0.252 & 0.748 \\
  36 &$^{ 80}$Kr & 1.30E+00 & 1.000 & 0.000 & 1.000 & 0.000 & 0.133 & 1.000 & 0.000 \\
  36 &$^{ 82}$Kr & 6.51E+00 & 1.000 & 0.000 & 1.000 & 0.000 & 0.274 & 1.000 & 0.000 \\
  36 &$^{ 83}$Kr & 6.45E+00 & 0.347 & 0.653 & 0.321 & 0.679 & 0.095 & 0.291 & 0.709 \\
  36 &$^{ 84}$Kr & 3.18E+01 & 0.370 & 0.630 & 0.257 & 0.743 & 0.117 & 0.222 & 0.778 \\
  36 &$^{ 86}$Kr & 9.61E+00 & 0.911 & 0.089 & 1.000 & 0.000 & 0.152 & 0.677 & 0.323 \\
  37 &$^{ 85}$Rb & 5.12E+00 & 0.198 & 0.802 & 0.440 & 0.560 & 0.153 & 0.326 & 0.674 \\
  37 &$^{ 87}$Rb & 2.11E+00 & 0.957 & 0.043 & 0.861 & 0.139 & 0.249 & 0.960 & 0.040 \\
  38 &$^{ 86}$Sr & 2.30E+00 & 1.000 & 0.000 & 1.000 & 0.000 & 0.611 & 1.000 & 0.000 \\
  38 &$^{ 87}$Sr & 1.60E+00 & 1.000 & 0.000 & 1.000 & 0.000 & 0.582 & 1.000 & 0.000 \\
  38 &$^{ 88}$Sr & 1.92E+01 & 0.869 & 0.131 & 0.787 & 0.213 & 0.711 & 0.903 & 0.097 \\
  39 &$^{ 89}$Y  & 4.63E+00 & 0.719 & 0.281 & 0.760 & 0.240 & 0.719 & 0.778 & 0.222 \\
  40 &$^{ 90}$Zr & 5.55E+00 & 0.821 & 0.179 & 0.529 & 0.471 & 0.603 & 0.807 & 0.193 \\
  40 &$^{ 91}$Zr & 1.21E+00 & 0.967 & 0.033 & 0.826 & 0.174 & 0.712 & 0.957 & 0.043 \\
  40 &$^{ 92}$Zr & 1.85E+00 & 0.705 & 0.295 & 0.966 & 0.034 & 0.682 & 0.745 & 0.255 \\
  40 &$^{ 94}$Zr & 1.87E+00 & 0.908 & 0.092 & 1.000 & 0.000 & 0.836 & 0.925 & 0.075 \\
  40 &$^{ 96}$Zr & 3.02E-01 & 0.000 & 1.000 & 1.000 & 0.000 & 0.387 & 0.291 & 0.709 \\
  41 &$^{ 93}$Nb & 7.80E-01 & 0.676 & 0.324 & 0.873 & 0.127 & 0.560 & 0.651 & 0.349 \\
  42 &$^{ 95}$Mo & 4.04E-01 & 0.470 & 0.530 & 0.653 & 0.347 & 0.454 & 0.479 & 0.521 \\
  42 &$^{ 96}$Mo & 4.25E-01 & 1.000 & 0.000 & 1.000 & 0.000 & 0.782 & 1.000 & 0.000 \\
  42 &$^{ 97}$Mo & 2.45E-01 & 0.642 & 0.358 & 0.670 & 0.330 & 0.433 & 0.563 & 0.437 \\
  42 &$^{ 98}$Mo & 6.22E-01 & 0.847 & 0.153 & 0.881 & 0.119 & 0.575 & 0.804 & 0.196 \\
  42 &$^{100}$Mo & 2.50E-01 & 0.000 & 1.000 & 0.096 & 0.904 & 0.023 & 0.038 & 0.962 \\
  44 &$^{ 99}$Ru & 2.27E-01 & 0.300 & 0.700 & 0.238 & 0.762 & 0.210 & 0.312 & 0.688 \\
  44 &$^{100}$Ru & 2.24E-01 & 1.000 & 0.000 & 1.000 & 0.000 & 0.801 & 1.000 & 0.000 \\
  44 &$^{101}$Ru & 3.04E-01 & 0.158 & 0.842 & 0.122 & 0.878 & 0.128 & 0.141 & 0.859 \\
  44 &$^{102}$Ru & 5.62E-01 & 0.444 & 0.556 & 0.440 & 0.560 & 0.433 & 0.476 & 0.524 \\
  44 &$^{104}$Ru & 3.32E-01 & 0.000 & 1.000 & 0.000 & 1.000 & 0.014 & 0.023 & 0.977 \\
  45 &$^{103}$Rh & 3.70E-01 & 0.160 & 0.840 & 0.197 & 0.803 & 0.118 & 0.122 & 0.878 \\
  46 &$^{104}$Pd & 1.51E-01 & 1.000 & 0.000 & 1.000 & 0.000 & 0.839 & 1.000 & 0.000 \\
  46 &$^{105}$Pd & 3.03E-01 & 0.129 & 0.871 & 0.123 & 0.877 & 0.107 & 0.129 & 0.871 \\
  46 &$^{106}$Pd & 3.71E-01 & 0.491 & 0.509 & 0.539 & 0.461 & 0.398 & 0.510 & 0.490 \\
  46 &$^{108}$Pd & 3.59E-01 & 0.609 & 0.391 & 0.669 & 0.331 & 0.505 & 0.625 & 0.375 \\
  46 &$^{110}$Pd & 1.59E-01 & 0.000 & 1.000 & 0.019 & 0.981 & 0.016 & 0.044 & 0.956 \\
  47 &$^{107}$Ag & 2.54E-01 & 0.195 & 0.805 & 0.169 & 0.831 & 0.001 & 0.134 & 0.866 \\
  47 &$^{109}$Ag & 2.36E-01 & 0.231 & 0.769 & 0.271 & 0.729 & 0.226 & 0.286 & 0.714 \\
  48 &$^{110}$Cd & 1.97E-01 & 1.000 & 0.000 & 1.000 & 0.000 & 0.776 & 1.000 & 0.000 \\
  48 &$^{111}$Cd & 2.01E-01 & 0.203 & 0.797 & 0.244 & 0.756 & 0.249 & 0.321 & 0.679 \\
  48 &$^{112}$Cd & 3.80E-01 & 0.503 & 0.497 & 0.537 & 0.463 & 0.499 & 0.575 & 0.425 \\
  48 &$^{113}$Cd & 1.92E-01 & 0.302 & 0.698 & 0.354 & 0.646 & 0.285 & 0.365 & 0.635 \\
  48 &$^{114}$Cd & 4.52E-01 & 0.672 & 0.328 & 0.619 & 0.381 & 0.591 & 0.693 & 0.307 \\
  48 &$^{116}$Cd & 1.18E-01 & 0.000 & 1.000 & 0.191 & 0.809 & 0.088 & 0.087 & 0.913 \\
  49 &$^{115}$In & 1.70E-01 & 0.320 & 0.680 & 0.347 & 0.653 & 0.298 & 0.391 & 0.609 \\
  50 &$^{116}$Sn & 5.24E-01 & 1.000 & 0.000 & 1.000 & 0.000 & 0.683 & 1.000 & 0.000 \\
  50 &$^{117}$Sn & 2.77E-01 & 0.533 & 0.467 & 0.458 & 0.542 & 0.384 & 0.501 & 0.499 \\
  50 &$^{118}$Sn & 8.73E-01 & 0.837 & 0.163 & 0.721 & 0.279 & 0.554 & 0.728 & 0.272 \\
  50 &$^{119}$Sn & 3.09E-01 & 0.705 & 0.295 & 0.405 & 0.595 & 0.469 & 0.618 & 0.382 \\
  50 &$^{120}$Sn & 1.18E+00 & 0.934 & 0.066 & 0.818 & 0.182 & 0.627 & 0.850 & 0.150 \\
  50 &$^{122}$Sn & 1.67E-01 & 0.000 & 1.000 & 0.090 & 0.910 & 0.364 & 0.101 & 0.899 \\
  50 &$^{124}$Sn & 2.09E-01 & 0.000 & 1.000 & 0.000 & 1.000 &  -    & 0.008 & 0.992 \\
  51 &$^{121}$Sb & 1.79E-01 & 0.294 & 0.706 & 0.533 & 0.467 & 0.309 & 0.393 & 0.607 \\
  51 &$^{123}$Sb & 1.34E-01 & 0.000 & 1.000 & 0.157 & 0.843 & 0.050 & 0.051 & 0.949 \\
  52 &$^{122}$Te & 1.22E-01 & 1.000 & 0.000 & 1.000 & 0.000 & 0.736 & 1.000 & 0.000 \\
  52 &$^{123}$Te & 4.30E-02 & 1.000 & 0.000 & 1.000 & 0.000 & 0.741 & 1.000 & 0.000 \\
  52 &$^{124}$Te & 2.26E-01 & 1.000 & 0.000 & 1.000 & 0.000 & 0.781 & 1.000 & 0.000 \\
  52 &$^{125}$Te & 3.35E-01 & 0.248 & 0.752 & 0.236 & 0.764 & 0.174 & 0.269 & 0.731 \\
  52 &$^{126}$Te & 8.89E-01 & 0.462 & 0.538 & 0.447 & 0.553 & 0.363 & 0.474 & 0.526 \\
  52 &$^{128}$Te & 1.49E+00 & 0.000 & 1.000 & 0.011 & 0.989 & 0.033 & 0.026 & 0.974 \\
  52 &$^{130}$Te & 1.58E+00 & 0.000 & 1.000 & 0.003 & 0.997 &    -  & 0.004 & 0.996 \\
  53 &$^{127}$I  & 1.10E+00 & 0.055 & 0.945 & 0.229 & 0.771 & 0.038 & 0.032 & 0.968 \\
  54 &$^{128}$Xe & 1.22E-01 & 1.000 & 0.000 & 1.000 & 0.000 & 0.763 & 1.000 & 0.000 \\
  54 &$^{129}$Xe & 1.50E+00 & 0.051 & 0.949 & 0.280 & 0.720 & 0.028 & 0.030 & 0.970 \\
  54 &$^{130}$Xe & 2.39E-01 & 1.000 & 0.000 & 1.000 & 0.000 & 0.851 & 1.000 & 0.000 \\
  54 &$^{131}$Xe & 1.19E+00 & 0.084 & 0.916 & 0.309 & 0.691 & 0.065 & 0.055 & 0.945 \\
  54 &$^{132}$Xe & 1.44E+00 & 0.384 & 0.616 & 0.546 & 0.454 & 0.268 & 0.353 & 0.647 \\
  54 &$^{134}$Xe & 5.27E-01 & 0.000 & 1.000 & 0.269 & 0.731 & 0.041 & 0.028 & 0.972 \\
  54 &$^{136}$Xe & 4.29E-01 & 0.000 & 1.000 & 0.231 & 0.769 &   -   & 0.001 & 0.999 \\
  55 &$^{133}$Cs & 3.71E-01 & 0.151 & 0.849 & 0.167 & 0.833 & 0.135 & 0.157 & 0.843 \\
  56 &$^{134}$Ba & 1.08E-01 & 1.000 & 0.000 & 1.000 & 0.000 & 1.011 & 1.000 & 0.000 \\
  56 &$^{135}$Ba & 2.95E-01 & 0.186 & 0.814 & 0.159 & 0.841 & 0.285 & 0.249 & 0.751 \\
  56 &$^{136}$Ba & 3.51E-01 & 1.000 & 0.000 & 1.000 & 0.000 & 1.013 & 1.000 & 0.000 \\
  56 &$^{137}$Ba & 5.02E-01 & 0.568 & 0.432 & 0.661 & 0.339 & 0.632 & 0.647 & 0.353 \\
  56 &$^{138}$Ba & 3.21E+00 & 0.940 & 0.060 & 0.933 & 0.067 & 0.918 & 0.959 & 0.041 \\
  57 &$^{139}$La & 4.57E-01 & 0.754 & 0.246 & 0.656 & 0.344 & 0.755 & 0.800 & 0.200 \\
  58 &$^{140}$Ce & 1.04E+00 & 0.909 & 0.091 & 0.846 & 0.154 & 0.920 & 0.935 & 0.065 \\
  58 &$^{142}$Ce & 1.31E-01 & 0.000 & 1.000 & 0.496 & 0.504 & 0.195 & 0.126 & 0.874 \\
  59 &$^{141}$Pr & 1.72E-01 & 0.491 & 0.509 & 0.360 & 0.640 & 0.499 & 0.535 & 0.465 \\
  60 &$^{142}$Nd & 2.31E-01 & 1.000 & 0.000 & 1.000 & 0.000 & 0.983 & 1.000 & 0.000 \\
  60 &$^{143}$Nd & 1.03E-01 & 0.363 & 0.637 & 0.315 & 0.685 & 0.329 & 0.377 & 0.623 \\
  60 &$^{144}$Nd & 2.03E-01 & 0.528 & 0.472 & 0.508 & 0.492 & 0.522 & 0.565 & 0.435 \\
  60 &$^{145}$Nd & 7.50E-02 & 0.290 & 0.710 & 0.280 & 0.720 & 0.262 & 0.297 & 0.703 \\
  60 &$^{146}$Nd & 1.47E-01 & 0.632 & 0.368 & 0.637 & 0.363 & 0.660 & 0.699 & 0.301 \\
  60 &$^{148}$Nd & 4.90E-02 & 0.083 & 0.917 & 0.141 & 0.859 & 0.173 & 0.156 & 0.844 \\
  60 &$^{150}$Nd & 4.80E-02 & 0.000 & 1.000 & 0.000 & 1.000 &    -  & 0.003 & 0.997 \\
  62 &$^{147}$Sm & 4.10E-02 & 0.088 & 0.912 & 0.185 & 0.815 & 0.265 & 0.234 & 0.766 \\
  62 &$^{148}$Sm & 3.00E-02 & 1.000 & 0.000 & 1.000 & 0.000 & 1.034 & 1.000 & 0.000 \\
  62 &$^{149}$Sm & 3.70E-02 & 0.139 & 0.861 & 0.127 & 0.873 & 0.129 & 0.121 & 0.879 \\
  62 &$^{150}$Sm & 2.00E-02 & 1.000 & 0.000 & 1.000 & 0.000 & 1.000 & 1.000 & 0.000 \\
  62 &$^{152}$Sm & 7.10E-02 & 0.254 & 0.746 & 0.196 & 0.804 & 0.228 & 0.268 & 0.732 \\
  62 &$^{154}$Sm & 6.00E-02 & 0.000 & 1.000 & 0.008 & 0.992 & 0.025 & 0.049 & 0.951 \\
  63 &$^{151}$Eu & 4.71E-02 & 0.000 & 1.000 & 0.040 & 0.960 & 0.059 & 0.047 & 0.953 \\
  63 &$^{153}$Eu & 5.14E-02 & 0.040 & 0.960 & 0.037 & 0.963 & 0.061 & 0.050 & 0.950 \\
  64 &$^{154}$Gd & 7.80E-03 & 1.000 & 0.000 & 1.000 & 0.000 & 0.891 & 1.000 & 0.000 \\
  64 &$^{155}$Gd & 5.33E-02 & 0.062 & 0.938 & 0.122 & 0.878 & 0.060 & 0.057 & 0.943 \\
  64 &$^{156}$Gd & 7.36E-02 & 0.214 & 0.786 & 0.213 & 0.787 & 0.179 & 0.206 & 0.794 \\
  64 &$^{157}$Gd & 5.63E-02 & 0.132 & 0.868 & 0.163 & 0.837 & 0.111 & 0.097 & 0.903 \\
  64 &$^{158}$Gd & 8.94E-02 & 0.318 & 0.682 & 0.313 & 0.687 & 0.275 & 0.308 & 0.692 \\
  64 &$^{160}$Gd & 7.87E-02 & 0.000 & 1.000 & 0.058 & 0.942 & 0.007 & 0.022 & 0.978 \\
  65 &$^{159}$Tb & 6.34E-02 & 0.063 & 0.938 & 0.052 & 0.948 & 0.080 & 0.072 & 0.928 \\
  66 &$^{160}$Dy & 9.40E-03 & 1.000 & 0.000 & 1.000 & 0.000 & 0.901 & 1.000 & 0.000 \\
  66 &$^{161}$Dy & 7.62E-02 & 0.051 & 0.949 & 0.028 & 0.972 & 0.053 & 0.046 & 0.954 \\
  66 &$^{162}$Dy & 1.03E-01 & 0.137 & 0.863 & 0.125 & 0.875 & 0.160 & 0.159 & 0.841 \\
  66 &$^{163}$Dy & 1.01E-01 & 0.021 & 0.979 & 0.033 & 0.967 & 0.044 & 0.046 & 0.954 \\
  66 &$^{164}$Dy & 1.14E-01 & 0.165 & 0.835 & 0.097 & 0.903 & 0.239 & 0.243 & 0.757 \\
  67 &$^{165}$Ho & 9.10E-02 & 0.067 & 0.933 & 0.078 & 0.922 & 0.083 & 0.074 & 0.926 \\
  68 &$^{166}$Er & 8.80E-02 & 0.143 & 0.857 & 0.144 & 0.856 & 0.167 & 0.134 & 0.866 \\
  68 &$^{167}$Er & 6.00E-02 & 0.086 & 0.914 & 0.090 & 0.910 & 0.094 & 0.088 & 0.912 \\
  68 &$^{168}$Er & 7.10E-02 & 0.299 & 0.701 & 0.287 & 0.713 & 0.314 & 0.367 & 0.633 \\
  68 &$^{170}$Er & 3.90E-02 & 0.026 & 0.974 & 0.054 & 0.946 & 0.126 & 0.126 & 0.874 \\
  69 &$^{169}$Tm & 4.06E-02 & 0.162 & 0.838 & 0.163 & 0.837 & 0.091 & 0.128 & 0.872 \\
  70 &$^{170}$Yb & 7.60E-03 & 1.000 & 0.000 & 1.000 & 0.000 & 0.958 & 1.000 & 0.000 \\
  70 &$^{171}$Yb & 3.61E-02 & 0.121 & 0.879 & 0.177 & 0.823 & 0.227 & 0.226 & 0.774 \\
  70 &$^{172}$Yb & 5.56E-02 & 0.333 & 0.667 & 0.315 & 0.685 & 0.439 & 0.454 & 0.546 \\
  70 &$^{173}$Yb & 4.13E-02 & 0.205 & 0.795 & 0.235 & 0.765 & 0.278 & 0.305 & 0.695 \\
  70 &$^{174}$Yb & 8.21E-02 & 0.519 & 0.481 & 0.524 & 0.476 & 0.609 & 0.635 & 0.365 \\
  70 &$^{176}$Yb & 3.33E-02 & 0.000 & 1.000 & 0.123 & 0.877 & 0.072 & 0.115 & 0.885 \\
  71 &$^{175}$Lu & 3.70E-02 & 0.162 & 0.838 & 0.176 & 0.824 & 0.181 & 0.176 & 0.824 \\
  71 &$^{176}$Lu & 1.10E-03 & 1.000 & 0.000 & 1.000 & 0.000 & 1.055 & 1.000 & 0.000 \\
  72 &$^{176}$Hf & 8.10E-03 & 1.000 & 0.000 & 1.000 & 0.000 & 1.001 & 1.000 & 0.000 \\
  72 &$^{177}$Hf & 3.16E-02 & 0.172 & 0.828 & 0.247 & 0.753 & 0.173 & 0.175 & 0.825 \\
  72 &$^{178}$Hf & 4.25E-02 & 0.488 & 0.512 & 0.548 & 0.452 & 0.575 & 0.618 & 0.382 \\
  72 &$^{179}$Hf & 2.12E-02 & 0.318 & 0.682 & 0.349 & 0.651 & 0.412 & 0.451 & 0.549 \\
  72 &$^{180}$Hf & 5.47E-02 & 0.636 & 0.364 & 0.735 & 0.265 & 0.894 & 0.809 & 0.191 \\
  73 &$^{181}$Ta & 2.10E-02 & 0.409 & 0.591 & 0.495 & 0.505 & 0.466 & 0.503 & 0.497 \\
  74 &$^{182}$W  & 3.63E-02 & 0.667 & 0.333 & 0.625 & 0.375 & 0.638 & 0.675 & 0.325 \\
  74 &$^{183}$W  & 1.96E-02 & 0.650 & 0.350 & 0.668 & 0.332 & 0.622 & 0.701 & 0.299 \\
  74 &$^{184}$W  & 4.20E-02 & 0.690 & 0.310 & 0.748 & 0.252 & 0.788 & 0.790 & 0.210 \\
  74 &$^{186}$W  & 3.90E-02 & 0.162 & 0.838 & 0.372 & 0.628 & 0.424 & 0.279 & 0.721 \\
  75 &$^{185}$Re & 2.07E-02 & 0.222 & 0.778 & 0.271 & 0.729 & 0.270 & 0.341 & 0.659 \\
  75 &$^{187}$Re & 3.74E-02 & 0.029 & 0.971 & 0.150 & 0.850 & 0.094 & 0.048 & 0.952 \\
  76 &$^{186}$Os & 1.08E-02 & 1.000 & 0.000 & 1.000 & 0.000 & 1.035 & 1.000 & 0.000 \\
  76 &$^{187}$Os & 8.60E-03 & 0.990 & 0.010 & 1.000 & 0.000 & 0.374 & 0.751 & 0.249 \\
  76 &$^{188}$Os & 9.04E-02 & 0.168 & 0.832 & 0.217 & 0.783 & 0.282 & 0.209 & 0.791 \\
  76 &$^{189}$Os & 1.10E-01 & 0.035 & 0.965 & 0.064 & 0.936 & 0.048 & 0.041 & 0.959 \\
  76 &$^{190}$Os & 1.79E-01 & 0.111 & 0.889 & 0.151 & 0.849 & 0.146 & 0.132 & 0.868 \\
  76 &$^{192}$Os & 2.78E-01 & 0.003 & 0.997 & 0.018 & 0.982 & 0.033 & 0.018 & 0.982 \\
  77 &$^{191}$Ir & 2.50E-01 & 0.020 & 0.980 & 0.084 & 0.916 & 0.019 & 0.015 & 0.985 \\
  77 &$^{193}$Ir & 4.21E-01 & 0.007 & 0.993 & 0.078 & 0.922 & 0.014 & 0.009 & 0.991 \\
  78 &$^{192}$Pt & 1.00E-02 & 1.000 & 0.000 & 1.000 & 0.000 & 0.812 & 1.000 & 0.000 \\
  78 &$^{194}$Pt & 4.20E-01 & 0.044 & 0.956 & 0.000 & 1.000 & 0.049 & 0.072 & 0.928 \\
  78 &$^{195}$Pt & 4.31E-01 & 0.013 & 0.987 & 0.000 & 1.000 & 0.020 & 0.032 & 0.968 \\
  78 &$^{196}$Pt & 3.22E-01 & 0.101 & 0.899 & 0.062 & 0.938 & 0.123 & 0.133 & 0.867 \\
  78 &$^{198}$Pt & 9.10E-02 & 0.000 & 1.000 & 0.000 & 1.000 & 0.001 & 0.027 & 0.973 \\
  79 &$^{197}$Au & 1.95E-01 & 0.054 & 0.946 & 0.021 & 0.979 & 0.061 & 0.058 & 0.942 \\
  80 &$^{198}$Hg & 4.60E-02 & 1.000 & 0.000 & 1.000 & 0.000 & 0.828 & 1.000 & 0.000 \\
  80 &$^{199}$Hg & 7.70E-02 & 0.271 & 0.729 & 0.342 & 0.658 & 0.216 & 0.255 & 0.745 \\
  80 &$^{200}$Hg & 1.06E-01 & 0.630 & 0.370 & 0.685 & 0.315 & 0.519 & 0.575 & 0.425 \\
  80 &$^{201}$Hg & 6.00E-02 & 0.426 & 0.574 & 0.558 & 0.442 & 0.399 & 0.428 & 0.572 \\
  80 &$^{202}$Hg & 1.37E-01 & 0.752 & 0.248 & 0.812 & 0.188 & 0.704 & 0.742 & 0.258 \\
  80 &$^{204}$Hg & 3.10E-02 & 0.000 & 1.000 & 0.142 & 0.858 & 0.082 & 0.038 & 0.962 \\
  81 &$^{203}$Tl & 5.40E-02 & 0.778 & 0.222 & 0.939 & 0.061 & 0.807 & 0.823 & 0.177 \\
  81 &$^{205}$Tl & 1.29E-01 & 0.594 & 0.406 & 0.615 & 0.385 & 0.667 & 0.735 & 0.265 \\
  82 &$^{204}$Pb & 6.60E-02 & 1.000 & 0.000 & 1.000 & 0.000 & 0.870 & 1.000 & 0.000 \\
  82 &$^{206}$Pb & 6.14E-01 & 0.594 & 0.406 & 0.679 & 0.321 & 0.729 & 0.688 & 0.312 \\
  82 &$^{207}$Pb & 6.80E-01 & 0.528 & 0.472 & 0.791 & 0.209 & 0.702 & 0.706 & 0.294 \\
  82 &$^{208}$Pb & 1.95E+00 & 0.931 & 0.069 & 1.000 & 0.000 & 0.977 & 0.922 & 0.078 \\
  83 &$^{209}$Bi & 1.38E-01 & 0.354 & 0.646 & 0.637 & 0.363 & 0.204 & 0.216 & 0.784 \\
  90 &$^{232}$Th & 4.40E-02 & 0.000 & 1.000 & 0.000 & 1.000 &   -   & 0.000 & 1.000 \\
  92 &$^{235}$U  & 5.80E-03 & 0.000 & 1.000 & 0.000 & 1.000 &   -   & 0.000 & 1.000 \\
  92 &$^{238}$U  & 1.80E-02 & 0.000 & 1.000 & 0.000 & 1.000 &   -   & 0.000 & 1.000 \\

\hline

\hline
\multicolumn{10}{|c|}{End of Table}\\
\hline\hline

\hline

\end{longtable}

\end{small}




\clearpage
\twocolumn

\begin{table}
\begin{center}
\caption{ Solar elementary abundances from \citet{Lod09} and fractions produced by the s-, r- and p- processes according to this study. }
\label{tab:Elements}
\begin{tabular}{c c c c c c c}
\hline
Elm. & Z & log $\epsilon(X)^{1}$  & Mass fraction & s- &  r-  & p- \\
\hline

Ga &   31 &    3.150 &   7.00E-08  &  0.730  &  0.270  &  0.000  \\
Ge &   32 &    3.645 &   2.28E-07  &  0.636  &  0.364  &  0.000  \\
As &   33 &    2.372 &   1.25E-08  &  0.581  &  0.419  &  0.000  \\
Se &   34 &    3.416 &   1.46E-07  &  0.379  &  0.612  &  0.008  \\
Br &   35 &    2.616 &   2.35E-08  &  0.265  &  0.735  &  0.000  \\
Kr &   36 &    3.334 &   1.28E-07  &  0.387  &  0.610  &  0.003  \\
Rb &   37 &    2.446 &   1.69E-08  &  0.510  &  0.490  &  0.000  \\
Sr &   38 &    2.953 &   5.58E-08  &  0.912  &  0.083  &  0.005  \\
Y  &   39 &    2.252 &   1.13E-08  &  0.778  &  0.222  &  0.000  \\
Zr &   40 &    2.619 &   2.70E-08  &  0.817  &  0.183  &  0.000  \\
Nb &   41 &    1.479 &   1.99E-09  &  0.651  &  0.349  &  0.000  \\
Mo &   42 &    1.993 &   6.71E-09  &  0.497  &  0.275  &  0.228  \\
Ru &   44 &    1.837 &   4.94E-09  &  0.338  &  0.591  &  0.071  \\
Rh &   45 &    1.155 &   1.04E-09  &  0.122  &  0.878  &  0.000  \\
Pd &   46 &    1.719 &   3.96E-09  &  0.448  &  0.542  &  0.010  \\
Ag &   47 &    1.277 &   1.45E-09  &  0.209  &  0.791  &  0.000  \\
Cd &   48 &    1.784 &   4.85E-09  &  0.548  &  0.432  &  0.021  \\
In &   49 &    0.837 &   5.60E-10  &  0.374  &  0.582  &  0.044  \\
Sn &   50 &    2.144 &   1.17E-08  &  0.680  &  0.301  &  0.019  \\
Sb &   51 &    1.082 &   1.05E-09  &  0.247  &  0.753  &  0.000  \\
Te &   52 &    2.258 &   1.64E-08  &  0.192  &  0.807  &  0.001  \\
I  &   53 &    1.628 &   3.83E-09  &  0.032  &  0.968  &  0.000  \\
Xe &   54 &    2.324 &   1.96E-08  &  0.182  &  0.816  &  0.002  \\
Cs &   55 &    1.156 &   1.35E-09  &  0.157  &  0.843  &  0.000  \\
Ba &   56 &    2.237 &   1.68E-08  &  0.888  &  0.109  &  0.002  \\
La &   57 &    1.247 &   1.74E-09  &  0.799  &  0.200  &  0.001  \\
Ce &   58 &    1.658 &   4.53E-09  &  0.848  &  0.148  &  0.004  \\
Pr &   59 &    0.822 &   6.64E-10  &  0.535  &  0.465  &  0.000  \\
Nd &   60 &    1.519 &   3.38E-09  &  0.615  &  0.385  &  0.000  \\
Sm &   62 &    1.013 &   1.10E-09  &  0.325  &  0.647  &  0.029  \\
Eu &   63 &    0.580 &   4.10E-10  &  0.049  &  0.951  &  0.000  \\
Gd &   64 &    1.143 &   1.55E-09  &  0.163  &  0.835  &  0.002  \\
Tb &   65 &    0.389 &   2.76E-10  &  0.072  &  0.928  &  0.000  \\
Dy &   66 &    1.193 &   1.80E-09  &  0.151  &  0.847  &  0.001  \\
Ho &   67 &    0.546 &   4.11E-10  &  0.074  &  0.926  &  0.000  \\
Er &   68 &    1.006 &   1.20E-09  &  0.184  &  0.799  &  0.017  \\
Tm &   69 &    0.195 &   1.88E-10  &  0.128  &  0.872  &  0.000  \\
Yb &   70 &    0.995 &   1.22E-09  &  0.429  &  0.570  &  0.001  \\
Lu &   71 &    0.168 &   1.83E-10  &  0.204  &  0.796  &  0.000  \\
Hf &   72 &    0.786 &   7.74E-10  &  0.605  &  0.393  &  0.002  \\
Ta &   73 &   -0.091 &   1.04E-10  &  0.503  &  0.497  &  0.000  \\
W  &   74 &    0.724 &   6.91E-10  &  0.601  &  0.397  &  0.001  \\
Re &   75 &    0.351 &   2.97E-10  &  0.154  &  0.846  &  0.000  \\
Os &   76 &    1.417 &   3.53E-09  &  0.103  &  0.897  &  0.000  \\
Ir &   77 &    1.413 &   3.53E-09  &  0.011  &  0.989  &  0.000  \\
Pt &   78 &    1.692 &   6.81E-09  &  0.078  &  0.922  &  0.000  \\
Au &   79 &    0.877 &   1.05E-09  &  0.058  &  0.942  &  0.000  \\
Hg &   80 &    1.248 &   2.52E-09  &  0.548  &  0.450  &  0.002  \\
Tl &   81 &    0.849 &   1.03E-09  &  0.760  &  0.240  &  0.000  \\
Pb &   82 &    2.106 &   1.88E-08  &  0.831  &  0.169  &  0.000  \\
Bi &   83 &    0.727 &   7.91E-10  &  0.216  &  0.784  &  0.000  \\
Th &   90 &    0.230 &   2.80E-10  &  0.000  &  1.000  &  0.000  \\
U  &   92 &   -0.037 &   1.55E-10  &  0.000  &  1.000  &  0.000  \\

\hline
\end{tabular}
\end{center}
Note:(1) log $\epsilon(X)\equiv$ log (X/H) +12.

\end{table}




\bibliographystyle{mnras}
\bibliography{Reference1} 






\bsp	
\label{lastpage}
\end{document}